\documentclass[12pt,letterpaper]{article}
\usepackage[markup=underlined]{changes}
\usepackage[a4paper, total={7in, 10in}]{geometry}

\usepackage{graphicx}
\usepackage{helvet}
\usepackage{authblk}
\usepackage{hyperref}
\usepackage{amsmath} 
\usepackage{amssymb} 
\usepackage{float} %
\usepackage{tikz}
\usepackage{pgfplots}
\pgfplotsset{compat=1.17}
\usepackage[super,comma,sort&compress]  
   {natbib}
\usepackage[right]{lineno} \linenumbers

\usepackage{subcaption} 
\usepackage{multirow}%

\makeatletter
\renewcommand{\maketitle}{\bgroup\setlength{\parindent}{0pt}
\begin{flushleft}
  \textbf{\@title}
  
  \@author
\end{flushleft}\egroup}
\makeatother

\title{Distribution Grids May Be a Barrier to Residential Electrification}

\date{}

\author[1]{Priyadarshan}
\author[3]{Constance Crozier}
\author[4,5]{Kyri Baker}
\author[1,2,6,*]{Kevin J. Kircher}

\affil[1]{Mechanical Engineering, Purdue University, West Lafayette, IN, USA}
\affil[2]{Electrical and Computer Engineering, Purdue University, West Lafayette, IN, USA}
\affil[3]{Industrial
and Systems Engineering, Georgia Institute of Technology, Atlanta, GA, USA}
\affil[4]{Civil, Environmental, and Architectural Engineering, University of Colorado Boulder, Boulder, CO, USA}
\affil[5]{Renewable and Sustainable Energy Institute, University of Colorado Boulder, Boulder, CO, USA}
\affil[6]{Lead contact}

\affil[*]{Correspondence: kircher@purdue.edu}

\begin{document}

\maketitle

\section*{SUMMARY}

Replacing fossil-fueled appliances and vehicles with electric alternatives can reduce greenhouse gas emissions and air pollution in many settings. However, electrification can also raise electricity demand beyond the safe limits of electrical infrastructure. This can increase the risk of blackouts or require grid reinforcement that is often slow and expensive. Here we estimate the physical and economic impacts on distribution grids of electrifying all housing and personal vehicles in each county of the lower 48 United States. We find that space heating is the main driver of grid impacts, with the coldest regions seeing demand peaks up to five times higher than today's peaks. Accommodating electrification of all housing and personal vehicles is estimated to require 600 GW of distribution grid reinforcement nationally, at a cost of \$350 to \$790 billion, or \$2,800 to \$6,400 per household (95\% confidence intervals). However, demand-side management could eliminate over two-thirds of grid reinforcement costs.

\section*{KEYWORDS}

Distribution grids, Electrification, Heat pumps, Electric vehicles

\section*{INTRODUCTION}

Fossil-fueled driving, space heating, and water heating account for about half of United States greenhouse gas emissions \cite{USEPA2023}. Replacing fossil-fueled vehicles and heaters with electric alternatives can significantly reduce these emissions today \cite{wilson2024heat}, and could essentially eliminate them in a future with 100\% carbon-free electricity. For these reasons, state and local governments have encouraged electrification through a range of policies, including subsidies for electric vehicles and appliances, as well as some bans on new fossil-fueled vehicles or appliances \cite{carb_2017}. Due in part to these policies, adoption of electric vehicles and appliances in the United States is accelerating \cite{nyserda_heat_pumps_2024}.

While electrification of fossil-fueled activities can deeply reduce emissions, it could also have serious impacts on the power grid. Today, electricity demand in most United States power systems peaks in the hottest hours of the year, driven mainly by air conditioning \cite{keskar2023planning}. Electrifying space heating can heighten annual demand peaks and shift them from summer to winter \cite{mai2018electrification,tarroja2018translating,amonkar2023differential,buonocore2022inefficient,bistline2021deep}. Electric vehicle adoption is likely to further exacerbate demand peaks \cite{ssembatya2024dual,protopapadaki2017heat,7275178}. Heightened demand peaks from electrification could cause power quality issues and make power outages more likely \cite{damianakis2023assessing,ssembatya2024dual}. Mitigating these risks often requires replacing power lines and transformers, which are typically sized to accommodate power flows in near-worst-case demand forecast scenarios. Transformers in particular are already in short supply, with United States utilities experiencing lead times of up to two years and prices that have increased by 400 to 900\% in the last three years \cite{mckenna2024major}. Utilities pass grid reinforcement costs on to ratepayers in the form of increased electricity prices \cite{larsen2018projecting,fant2020climate}, which weaken economic incentives for further electrification. While increased electricity sales from electrification of heating and other end uses may cover some of these costs, and can even put downward pressure on retail electricity prices\cite{synapseEVsDrivingRatesDown2019}, the overall economic implications of distribution grid reinforcement under deep electrification remain understudied.

Past research on the grid impacts of electrification focused mainly on high-voltage transmission systems. At the nationwide scale, the National Renewable Energy Laboratory's {\it Electrification Futures Study} \cite{steinberg2017electrification, mai2018electrification, murphy2021electrification} investigated how electrification might alter load profiles and transmission needs. In their high-electrification scenario (about 90\% adoption of electric vehicles and 70\% of heat pumps), peak electricity demand rose by 38\%. A nationwide study from the Electric Power Research Institute \cite{epri} estimated that 75\% adoption of electric vehicles and 50\% of heat pumps (by floor area) would raise peak demand by 24\% to 52\%. Wilson et al. \cite{wilson2024heat} used the ResStock tool \cite{reyna2022us} to investigate nationwide cost and emission impacts under various heat pump adoption scenarios. While Wilson et al. \cite{wilson2024heat} did not explicitly consider grid impacts, they stressed the need for future work to understand how evolving load profiles should shape infrastructure investments. At the regional scale, an American Council for an Energy-Efficient Economy \cite{Specian2021} study found that 30\% adoption of EVs and 100\% of heat pumps could increase peak demand in New England by 235\%. White et al. \cite{white2021quantifying} estimated that electrifying 100\% of space heating would require 25\% expansion of Texas grid capacity. Zhang et al. \cite{zhang2020quantifying} estimated that 10\% adoption of electric vehicles with uncoordinated charging would increase Midwest peak demand by 10\%.

In parallel with high-voltage transmission grid research at the nationwide and regional scales, several studies investigated medium-voltage distribution grid impacts in specific states or for vehicle electrification only. Elmallah et al. \cite{elmallah2022can} used the Distribution Deferral Opportunity Report, which the California Public Utilities Commission requires utilities to file each year, to analyze electrification impacts within the Pacific Gas \& Electric service territory. Elmallah et al. estimated that distribution grid reinforcement to accommodate new load from electrification, including up to 100\% adoption of electric vehicles and up to 50\% adoption of heat pumps and water heaters, could cost \$1 to \$10 billion (about \$200 to \$2,000 per household). Li et al. \cite{li2024impact} used the same dataset to study 100\% adoption of electric vehicles (but not heat pumps or water heaters) in California, estimating that 67\% of distribution feeders could need reinforcement, at a total cost of \$6 to \$20 billion (about \$300 to \$900 per household). Mai et al. estimated statewide distribution grid reinforcement costs in New York with up to 100\% electric vehicle adoption at \$1.4 to \$26.8 billion (about \$200 to \$3,500 per household) in their lowest- to highest-impact scenarios  \cite{nyserda2022_tedi}. Cutter et al. estimated that 100\% electric vehicle adoption across the United States could require up to \$200 billion in distribution grid reinforcement (about \$1,500 per household) \cite{ethree2021_distribution_costs}.Several other state-specific studies investigated the impacts of electrification on greenhouse gas emissions without considering impacts on power grids \cite{bistline2021deep, ebrahimi2018California, wei2013deep}.

\begin{figure}
\centering
\includegraphics[ clip,width = \linewidth]{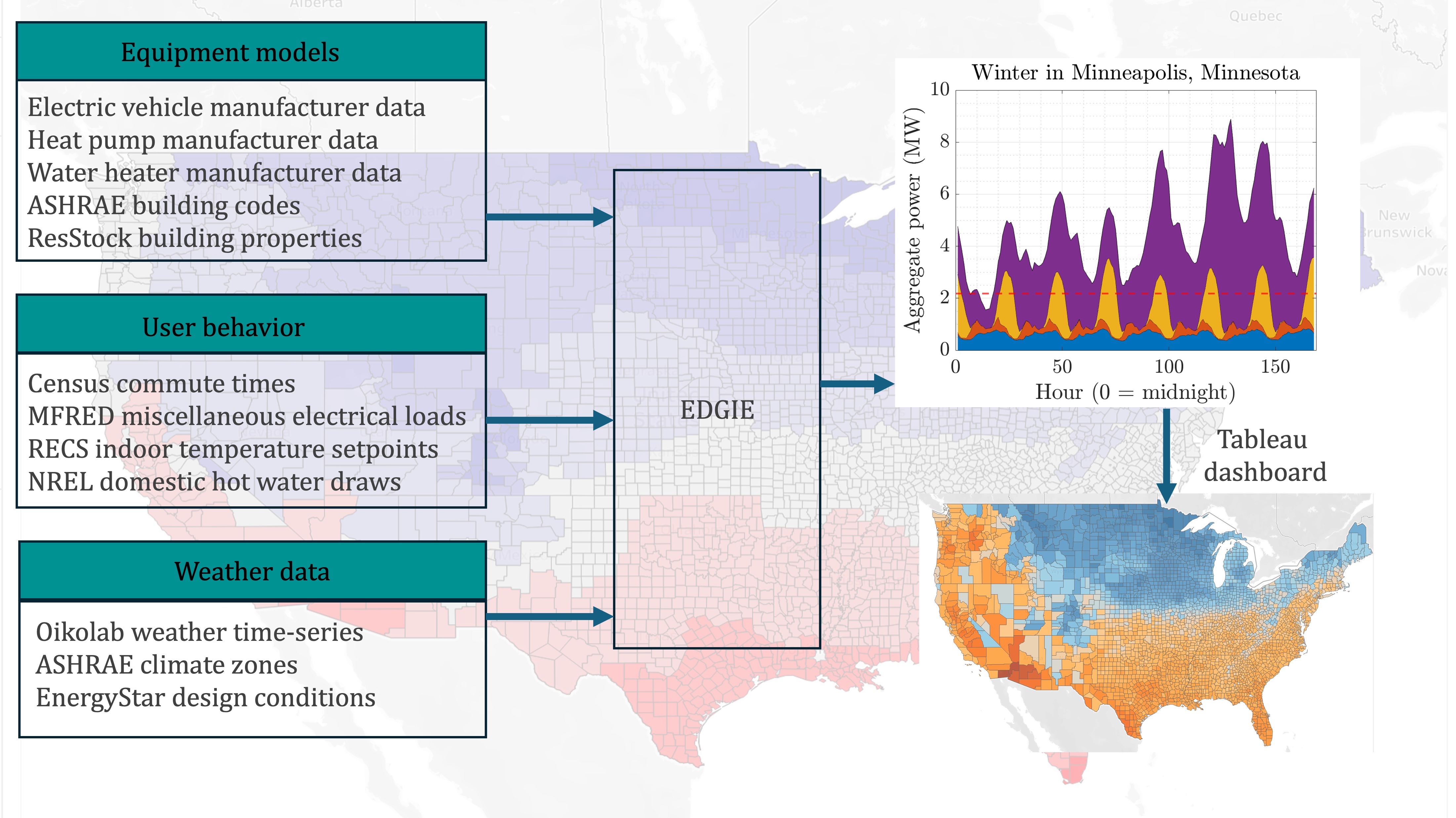}
\caption{The open-source EDGIE toolbox uses high-resolution, bottom-up modeling to estimate the impacts of home and vehicle electrification on distribution grids in each county of the lower 48 United States.}\label{informationFlow}
\end{figure}

As a complement to nationwide and regional transmission-level studies and California-specific distribution-level studies, this paper investigates the impacts of electrification on distribution grids nationwide. In addition to expanding the scope of distribution-level studies from the 58 counties in California to 3,100 counties in the lower 48 United States, this paper investigates strategies to mitigate grid impacts through demand-side management of space heating, water heating, and personal vehicles. These are the three main drivers of grid impacts from residential electrification, and we are aware of no distribution-level study that has considered demand-side management strategies for all of them. We find that almost all United States counties could require grid reinforcement in a 100\% electrification scenario. On average over the three coldest climate zones (zones 5, 6, and 7), future peak demand is three times higher than today's peak. Distribution grid reinforcement to accommodate 100\% electrification nationwide could cost \$350 to \$790 billion, or about \$2,800 to \$6,400 per household. However, the demand-side management strategies evaluated here, such as improving equipment efficiencies and coordinating device operation, could reduce nationwide grid reinforcement costs by up to 71\%.

We estimate distribution grid reinforcement requirements through bottom-up simulation. We estimate today's distribution grid capacity in each county by simulating all the appliances and vehicles in 1,000 representative housing units during peak heating and cooling weeks, rescaling the overall peak aggregate demand to reflect the true number of housing units, and adding a safety margin to represent headroom in today's capacity. We repeat the same process for an all-electric future. The estimated grid reinforcement requirement is the difference between the all-electric future grid capacity and today's grid capacity. To simulate buildings and vehicles, we use the open-source EDGIE (Emulating the Distribution Grid Impacts of Electrification) toolbox \cite{priyadarshan2024edgie}. As illustrated in Fig. \ref{informationFlow}, EDGIE models heat pumps, electric vehicles, water heaters, and building thermal dynamics. It uses real data on weather \cite{oiko}, indoor temperature setpoints \cite{RECS}, domestic hot water use \cite{hpwh}, and miscellaneous electrical loads \cite{meinrenken2020mfred}. EDGIE implements empirically validated, multi-physics modeling at hourly or sub-hourly time resolution and county-level spatial resolution. We tune building, appliance, and vehicle parameters to county-specific data. We make all data publicly available to facilitate reproduction or extension of the work.

\section*{RESULTS}

\subsection*{Impacts on physical infrastructure}
\label{Impacts on physical infrastructure}

\begin{figure}
  \centering
  \begin{subfigure}{0.5\textwidth}
    \centering
    \includegraphics[trim=0.2cm 0 0 0,clip,width=0.9\linewidth]{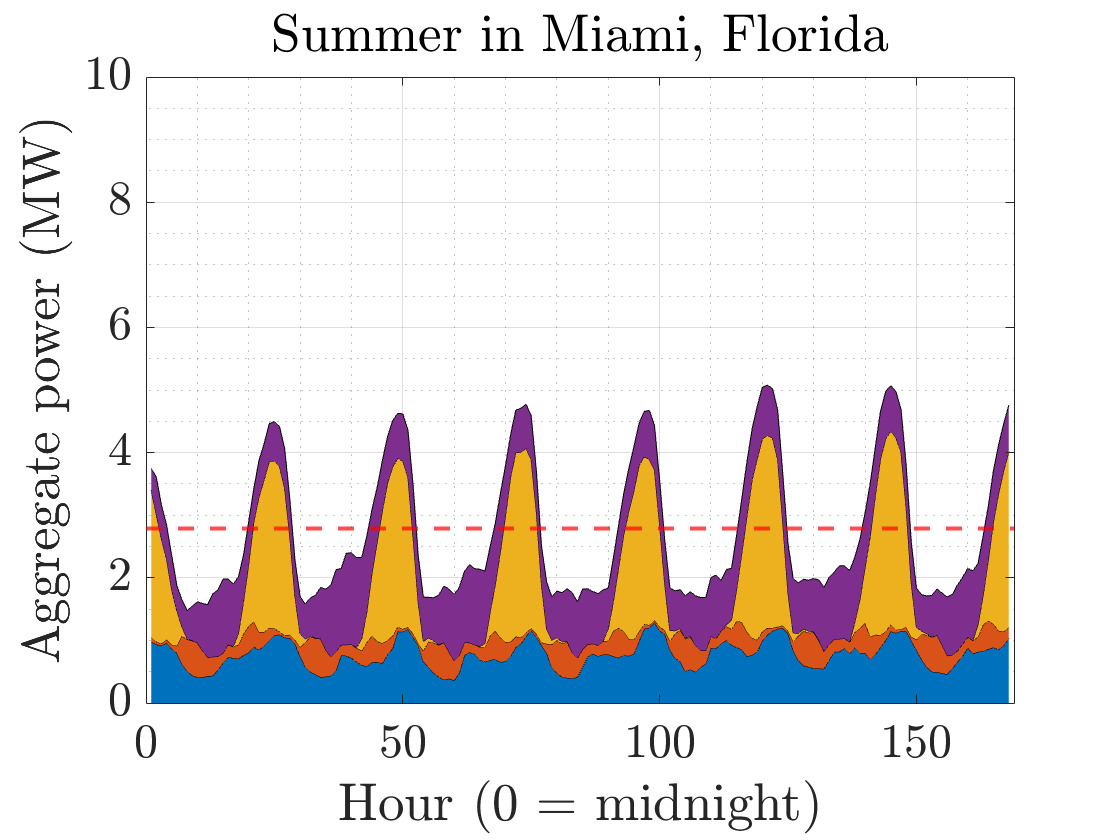}
    \label{fig:miami}
  \end{subfigure}%
  \begin{subfigure}{0.5\textwidth}
    \centering
    \includegraphics[trim=0cm 0 0 0,clip,width=0.9\linewidth]{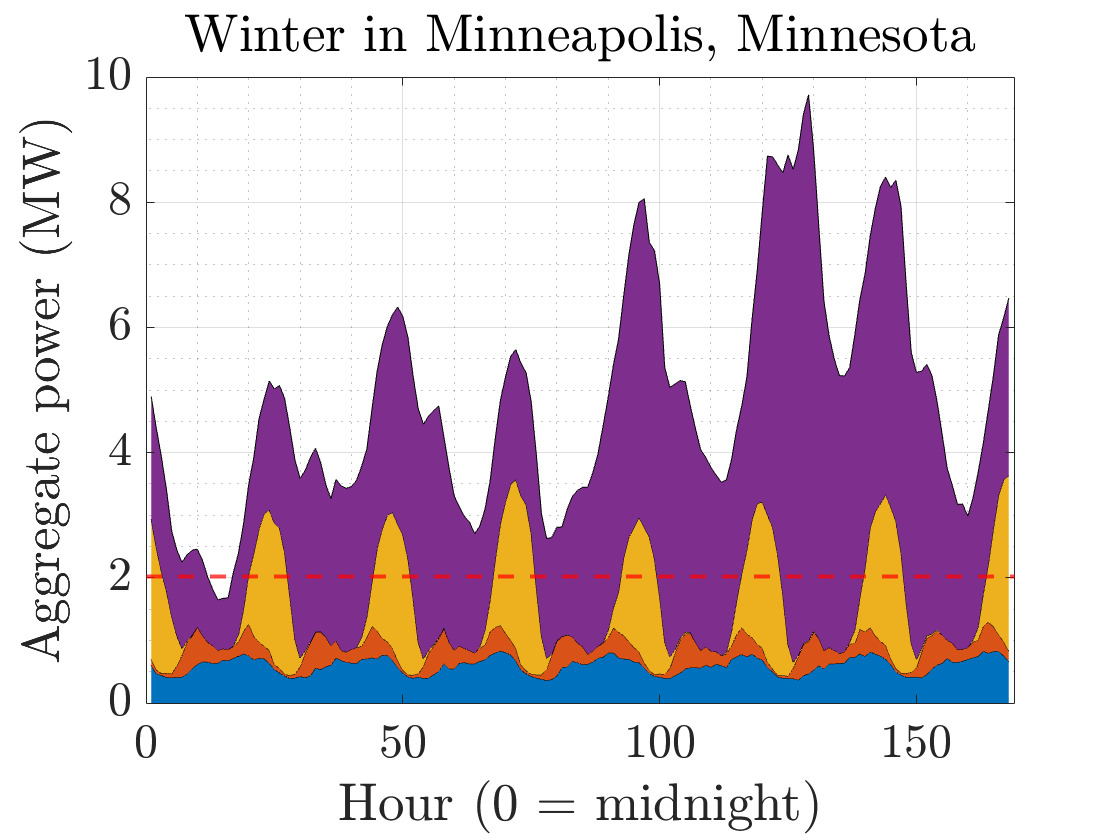}
    \label{fig:minnesota}
  \end{subfigure}

  \begin{subfigure}{\textwidth}
    \centering
    \includegraphics[width=0.8\linewidth]{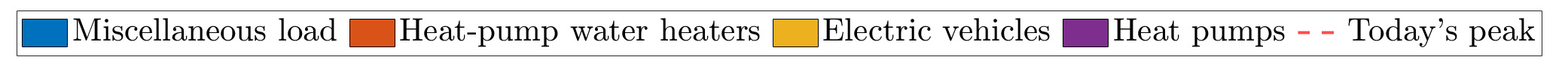}
    \label{fig:legends}
  \end{subfigure}

  \caption{In a hot week in Miami (left) and a cold week in Minneapolis (right), electricity demand peaks from 1,000 households exceed today's peaks factors of two and four, respectively.}
  \label{loadprofile}
\end{figure}

We assess the physical impacts of home and vehicle electrification on distribution grids through bottom-up modeling. In each county, we simulate all the residential appliances and personal vehicles from 1,000 representative households, then rescale the aggregate load profile to reflect the true number of households. Simulated appliance and vehicle properties, housing sizes and types, and levels of insulation and air sealing all vary both across and within counties. We fit physical and behavioral parameters associated with buildings, thermal equipment, and vehicles to county- and climate-specific data, using real data wherever possible. The Methods section and the Supplemental Notes provide more details on models and data sources (Fig. S1-S7).

We simulate peak heating and cooling weeks under two scenarios: (1) a business-as-usual scenario that reflects today's building stock, appliances, and vehicles; and (2) an all-electric scenario that reflects a possible future with complete electrification of personal vehicles and residential space and water heating. In the all-electric scenario, each home has a central air-to-air heat pump with backup resistance heat. We compute design heating and cooling loads for each home under the 99\% heating and 1\% cooling design temperatures from the American Society of Heating, Refrigeration, and Air-Conditioning Engineers {\it 2017 Handbook of Fundamentals} and the Air Conditioning Contractors of America {\it Manual J Design Conditions 8th Edition}. Each central heat pump is sized to meet the larger of the home's design heating load and design cooling load, up to a cap based on today's equipment sizes. We model the largest available central heat pump as having five tons (17.6 kW) of nameplate cooling capacity. If the design heating or cooling load exceeds the maximum capacity of the largest available central heat pump under design conditions, we add a mini-split heat pump with one to three tons (3.5 to 10.6 kW) of nameplate cooling capacity. Resistance backup meets any heat demand in excess of the combined capacities of the central heat pump and the mini-split heat pump. In the two northernmost climate zones, we simulate adoption of heat pumps whose coefficients of performance meet the United States Department of Energy's Cold Climate Heat Pump Challenge specifications \cite{doe2024cold}. The modeled heat pump capacities and coefficients of performance vary with outdoor temperatures.

\begin{figure}
\centering
\includegraphics[  trim=0cm 0cm 0cm 0cm,clip,width = \linewidth]{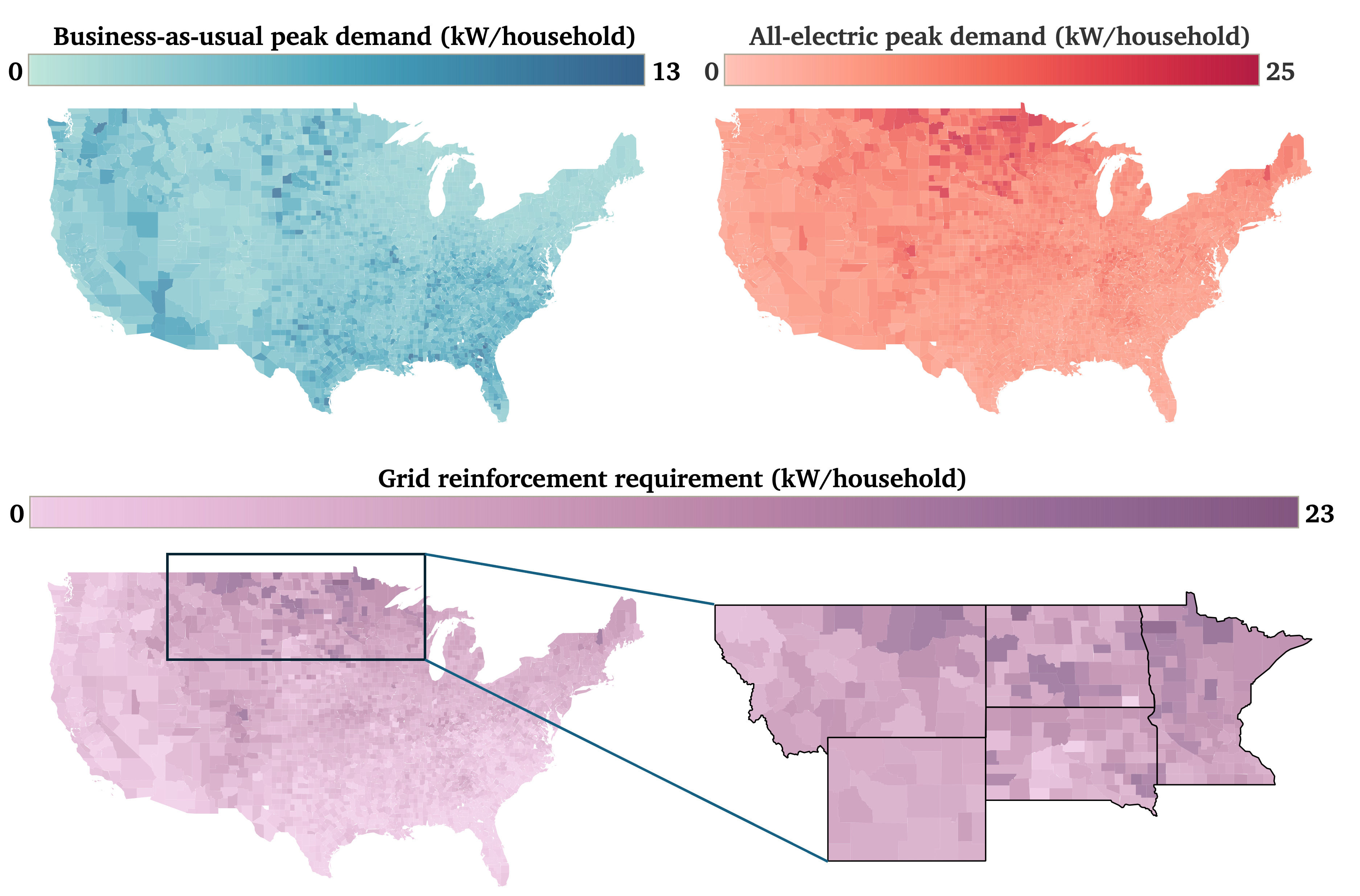}
\caption{Today (top left), per-household demand peaks are higher in hotter areas and in cold rural areas with significant amounts of electric heating. In an all-electric future (top right), per-household peaks increase and shift to colder areas. Per-household grid reinforcement requirements (bottom; the difference between the future peak and today's peak, adjusted for distribution grid headroom) are highest in the Rocky Mountains, the upper Midwest, and the Northeast. Each county's estimates are computed by simulating all devices in 1,000 representative households, then rescaling to the true number of households.
}\label{todayFuture}
\end{figure}

The modeling approach outlined above enables analysis of the sizes and causes of likely grid impacts from electrification. Fig.  \ref{loadprofile} shows the aggregate power used by 1,000 households over a peak cooling week in Miami, Florida (left plot; a hot, humid climate) and a peak heating week in Minneapolis, Minnesota (right; cold). In these plots, the shaded areas from bottom to top represent miscellaneous electrical loads (blue; lights, computers, dishwashers, etc.), water heaters (orange), electric vehicles (yellow), and air-source heat pumps with backup resistance heat (purple). Dashed red lines show today's annual demand peaks. In Miami, electrification doubles peak demand relative to today's peak, due mainly to electric vehicle charging. In Minneapolis, peak demand increases by a factor of five, driven mainly by space heating in the coldest weather, when heat demand rises and heating efficiencies fall.

We analyze grid impacts nationwide by running simulations for each county in the lower 48 United States. Fig. \ref{todayFuture} shows heat maps of the estimated peak demand, normalized by the number of homes, in each county under the business-as-usual scenario (top left) and the all-electric scenario (top right). The top left plot shows that normalized peak demand (in units of kW per household) in the business-as-usual scenario tends to be higher in areas with more cooling demand, reaching a maximum of 13 kW per household in the hot, sunny Southwest. By contrast, the top right plot shows that normalized peak demand in the all-electric scenario is higher in colder areas, reaching a maximum of 25 kW per household in the northern Midwest.

We estimate the distribution grid reinforcement requirement as the difference between the all-electric and business-as-usual distribution grid capacities. We estimate the distribution grid capacities in each county and scenario by simulating the appliances and vehicles for 1,000 households, rescaling the aggregate demand to the true number of households, taking the 99th percentile, then adding a safety margin to reflect distribution grid headroom. We base headroom estimates on a hosting capacity analysis for seven utilities in New York State \cite{Takahashi2024}, which found that today's headroom varies by utility from 15\% to 36\%. We use this headroom range for the business-as-usual scenario, randomizing over counties. For the all-electric scenario, we assume all counties have 20\% headroom. The discussion section analyzes the sensitivity of grid reinforcement requirements to headroom assumptions in the business-as-usual and all-electric scenarios.

The lower heat map in Fig. \ref{todayFuture} shows the estimated grid reinforcement requirement, normalized by the number of households, required to accommodate 100\% electrification in each county. The northern Midwest, Northeast, and Rocky Mountain regions require more grid reinforcement due to higher space heating demand. Grid reinforcement requirements reach a maximum of 23 kW per household in the northern Midwest. Many southern areas with mild winters require little or no grid reinforcement. In some cases, the lower heat map in Fig. \ref{todayFuture} shows significant variation in grid reinforcement requirements between adjacent counties. This variation can be explained by differences in housing types and sizes, levels of insulation and air sealing, driving patterns, weather extremes, or other location-specific data. These geographic differences suggest a need for tailored infrastructure planning, as approaches that work for one county may not necessarily work for neighboring counties.

\subsection*{Distribution grid reinforcement costs}
\label{Distribution grid reinforcement costs}

To estimate the economic costs of reinforcing distribution grids to accommodate electrification, we multiply the aggregate grid reinforcement requirement for each county (in units of kW) by a \$/kW price. In reality, grid reinforcement prices vary significantly across the United States due to variation in grid topology and capacity, prevalence of overhead vs. underground power lines, costs of labor and equipment, and other factors. Due to a lack of reliable location-specific data on grid reinforcement prices, however, we use the same price distribution for each county.

\begin{figure}
\centering
\includegraphics[ clip,width = \linewidth]{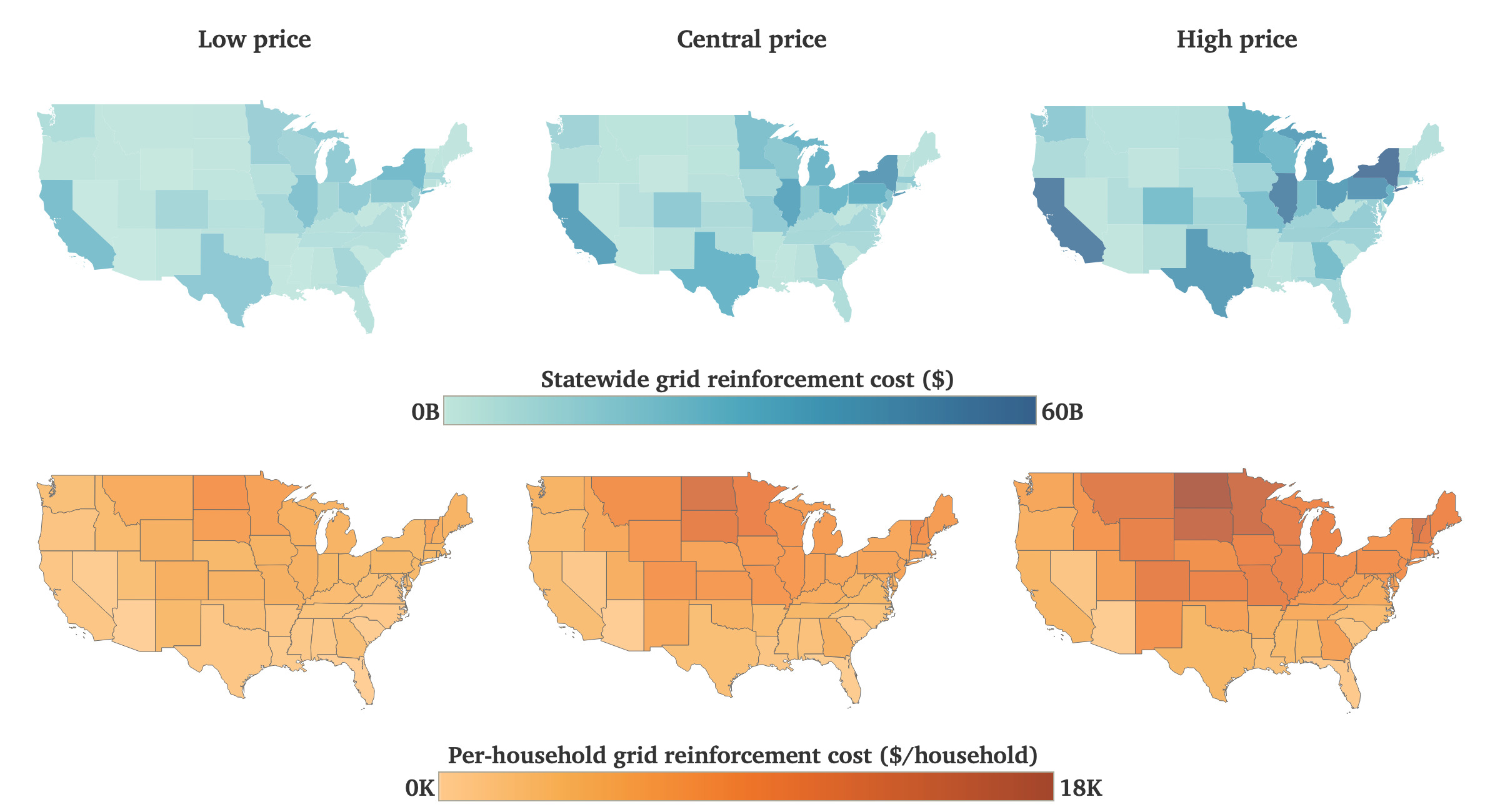}
\caption{Statewide distribution grid reinforcement costs (top row) reflect both population densities and per-household grid reinforcement requirements. Per-household costs (bottom row) do not reflect population densities. Left column: Low-price scenario for grid reinforcement. Middle: Central price. Right: High price.}\label{CostSensitivity}
\end{figure}

We draw on three studies \cite{elmallah2022can,zhang2020princeton,rauschkolb2021estimating} to determine a plausible price range for typical distribution grid reinforcement, which may include installing power lines, transformers, switchgear, capacity banks, or other infrastructure. Based on Pacific Gas \& Electric data, Elmallah et al. \cite{elmallah2022can} estimated prices of 860 to 36,500 \$/kW, with the highest prices for the smallest projects. (We inflation-adjust all price data to 2024 dollars.) Zhang et al. \cite{zhang2020princeton} used Energy Information Administration data \cite{EIA_AEO2019} to derive state-wise prices and a national average of 1,664 \$/kW, including a variety of administrative costs. Rauschkolb et al. \cite{rauschkolb2021estimating} analyzed data from the Federal Energy Regulatory Commission's Form 1 \cite{oedi_489}, where utilities can report their capital, operation, and maintenance costs related to grid reinforcement. By contrast to prior studies, Raushkolb et al. separated the regular costs of sustaining an existing distribution grid from the incremental costs of expanding grid capacity to accommodate load growth. Rauschkolb et al. developed a linear model that accounts for both one-time capital costs from expanding grid capacity, and for increased annual costs from operating and maintaining a larger grid.

Here, we modify Rauschkolb et al.'s approach to estimate a representative price range for distribution grid reinforcement. We assume grid capacities increase linearly from the business-as-usual capacities in 2025 to the all-electric capacities in 2050. We map future costs into net present costs using discounted cash flow analysis. Similar to Rauschkolb et al.'s approach, our modified approach accounts for both one-time capital costs of reinforcing infrastructure, and increased annual costs from operating and maintaining a larger grid. This approach captures distribution grid reinforcement costs incurred at primary components, such as substation transformers and high-capacity feeders, as well as secondary components, such as final line transformers and lower-capacity lines. Our approach implicitly assumes that reinforcement requirements for both primary and secondary components scale with the increase in system-wide peak demand. This assumption may lead to overestimating the costs of reinforcing secondary components, which experience demand peaks that may not coincide with system-wide peaks. However, this overestimation is likely small, as the \$/kW price of reinforcing secondary components is about two orders of magnitude lower than the \$/kW price of reinforcing primary components\cite{E3_DER_Grid_Impacts}. We do not model electrical infrastructure costs typically borne by individual customers, such as replacing service drops, circuit breaker panels, or wiring within buildings \cite{pergantis2025protecting}. The Methods section contains the details of our modifications, which yield a mean grid reinforcement price of 960 \$/kW and a 95\% confidence interval of 587 to 1,331 \$/kW. 

The cost of expanding distribution capacity can vary depending on whether it involves extending infrastructure to new construction  vs. retrofitting existing infrastructure to serve higher demand from the same customers. For instance, replacing an aging transformer with a higher-capacity transformer during routine maintenance, or upsizing a line during undergrounding for risk mitigation, may add significant capacity at a comparatively low incremental cost. The Raushkolb et al. model used here was fit to cost data from a period where distribution grid expansion was driven mainly by new construction, rather than retrofitting existing infrastructure. As grid reinforcement to accommodate electrification will likely involve substantial retrofitting of existing infrastructure, our cost estimates may be somewhat high.

The upper row of heat maps in Fig. \ref{CostSensitivity} shows the estimated total distribution grid reinforcement cost for each of the lower 48 United States under low (left), central (middle), and high (right) price scenarios. We compute these estimates by multiplying the grid reinforcement requirement, aggregated over all counties in the state, by prices of 587, 960, and 1,331 \$/kW, corresponding to the 2.5th, 50th, and 97.5th price percentiles. Unlike the per-household grid reinforcement requirement estimates in Fig. \ref{todayFuture}, the aggregate cost estimates in Fig. \ref{CostSensitivity} reflect population density as well as load growth intensity. States with both high populations and high per-household grid reinforcement requirements have the highest total costs. For example, the 7.6 million households in New York, a densely populated state with cold winters, see a total cost of about \$60 billion, or about \$7,900 per household. Per-household grid reinforcement costs, shown in the bottom row of heat maps in Fig. \ref{CostSensitivity}, reach a maximum of \$18,000 per household under the 97.5th price percentile in the cold, sparsely populated state of North Dakota.

The grid reinforcement cost estimates described above represent incremental spending, above and beyond the business-as-usual spending required to sustain existing distribution grids. The nationwide grid reinforcement cost estimate over 25 years in the central price scenario is \$570 billion. For comparison, we analyzed distribution grid spending data from the Federal Energy Regulatory Commission's Form 1 \cite{oedi_489} from 1995 to 2019. Over that 25-year period, the total inflation-adjusted spending on distribution grid infrastructure -- including all capital, operation, and maintenance costs associated with both sustaining existing grids and expanding capacity to accommodate load growth from residential, commercial, and industrial customers -- was \$975 billion. Relative to this cost, the estimated incremental grid reinforcement cost of \$570 billion to accommodate home and vehicle electrification is a 58\% increase. While this is not an exact comparison, it gives an approximate sense of scale for the incremental distribution grid spending required to accommodate residential electrification.

\vspace{1em}

\noindent \fbox{\begin{minipage}{\textwidth}
Aggregated over the lower 48 United States, distribution grid reinforcement to accommodate 100\% electrification of homes and private vehicles over 25 years is estimated to cost an additional \$350 to \$790 billion (\$110 to \$250 per household per year), above and beyond business-as-usual spending. For context, United States utilities spent \$975 billion on distribution grid infrastructure in the 25 years from 1995 to 2019 (\$390 per customer per year) \cite{oedi_489}. Residential energy bills cost Americans about \$230 billion annually (\$1,800 per household per year)  \cite{eia_table_ce11_2020}. The Inflation Reduction Act of 2022 allocated about \$370 billion over ten years (\$290 per household per year) to spending on energy and climate \cite{technologyreview2022climate}. The United States annual military budget was about \$820 billion (\$6,500 per household per year) in 2023 \cite{dod2022defense}.
\end{minipage}}

\subsection*{Cost reductions from smart electrification}

\begin{figure}
\centering
\includegraphics[ clip,width = \linewidth]{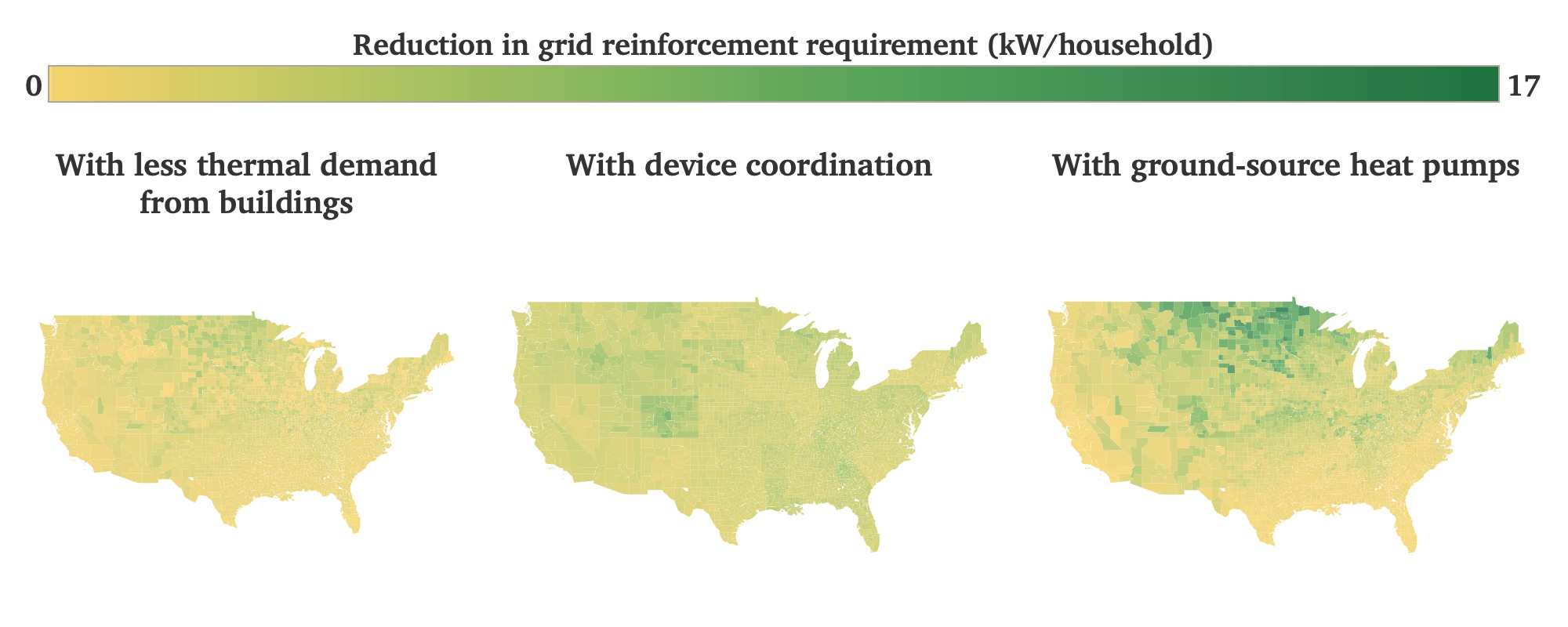}
\caption{Decreasing thermal demand from buildings (left) and switching from air- to ground-source heat pumps (right) reduce grid reinforcement requirements, particularly in cold climates. Coordinating device operation (middle) reduces grid reinforcement requirements in all climates.}\label{strategies}
\end{figure}

The estimated distribution grid reinforcement requirements and costs in the previous sections come from an electrification scenario that extrapolates forward today's housing stock, vehicle fleet, equipment configurations and efficiencies, and user behavior. Unmodified, these electrification choice are estimated to require \$350 to \$790 billion in distribution grid reinforcement costs nationwide. However, electrification with strategic demand-side management -- referred to here as `smart electrification' -- could reduce these costs substantially.

To assess the cost reduction potential of smart electrification, we simulate three demand-side management strategies. First, thermal demand from buildings could decrease due to better insulation and air sealing, due to shifts from detached housing to less energy-intensive attached housing, and/or due to shifts to smaller housing. We simulate thermal demand reduction by increasing the effective thermal resistance (which models the combined effects of wall and roof insulation, window quality, outdoor air infiltration rates, and the surface area exposed to the outdoor air) by 25\%, consistent with the `basic enclosure package' from Maxim and Grubert \cite{maxim2023highly}. Second, the air-source heat pumps simulated above could shift to ground-source heat pumps, which use about one-third less electricity per unit of heat output and require less backup resistance heat. Third, software could reduce peak aggregate demand by coordinating the operation of water heaters, electric vehicles, and heat pumps. For example, heat pumps could preheat homes in anticipation of aggregate demand peaks, or electric vehicles could interleave their charging with heat pump operation. (We simulate device coordination through convex optimization with perfect information in one representative location per state, as discussed in the Supplemental Notes.) The demand-side management strategies simulated here are far from exhaustive. We leave for future work a range of measures, such as solar photovoltaics, bidirectional electric vehicle charging, home batteries, and thermal storage, that could further reduce grid reinforcement requirements.

\begin{figure}
\centering
\includegraphics[ clip,width = \linewidth]{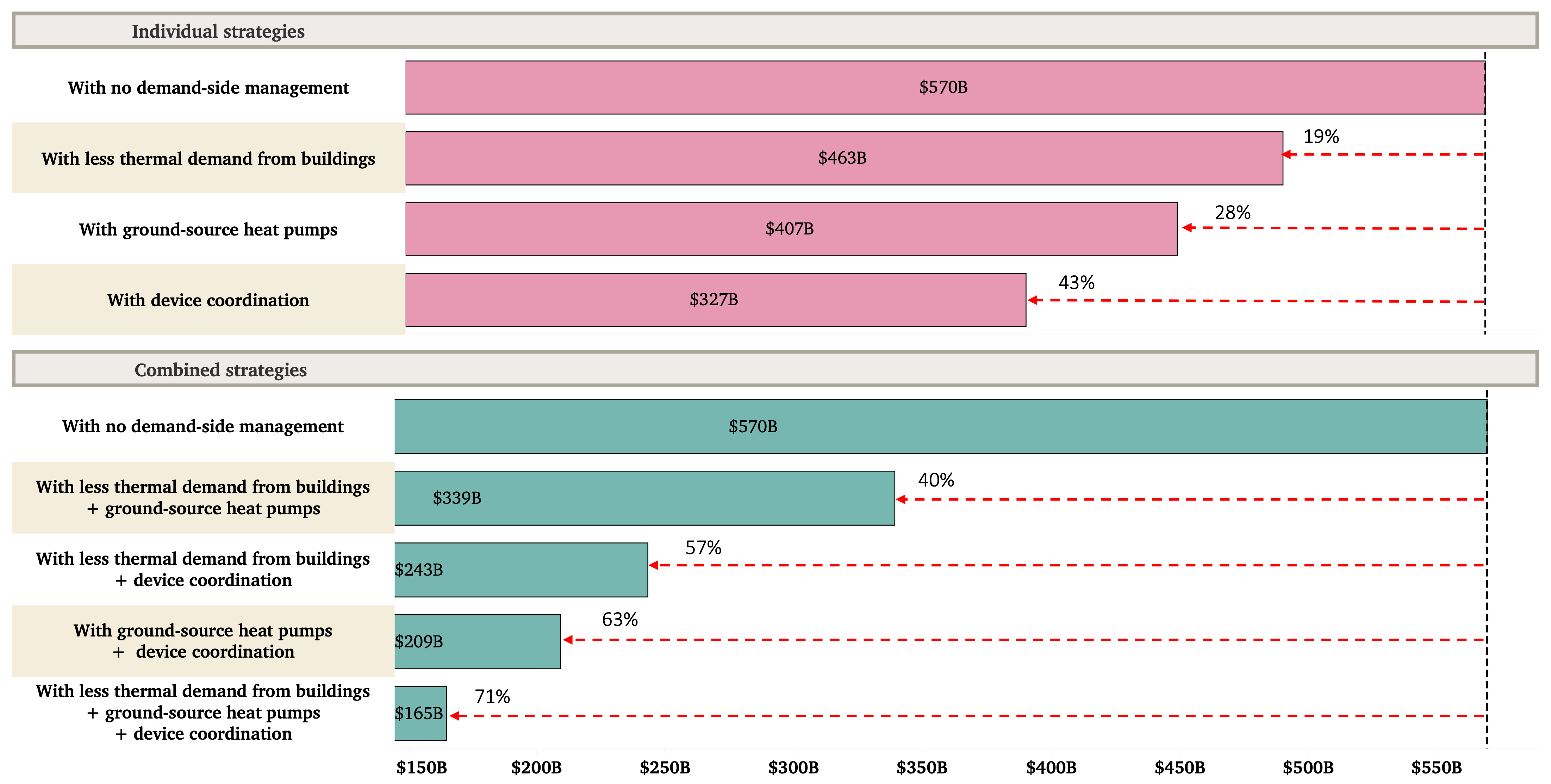}
\caption{Grid reinforcement cost reductions from individual (top set of bars) and combined (bottom) demand-side management strategies.}\label{barGrapgh}
\end{figure}

Fig. \ref{strategies} shows heat maps of the estimated reductions in grid reinforcement requirements in each county under the three demand-side management strategies considered here. The heat maps with less thermal demand from buildings (left) in Fig. \ref{strategies} and with ground-source heat pumps (right) show significant reductions in grid reinforcement requirements, particularly in the coldest regions. The heat map with device coordination (middle) shows modest reductions in grid reinforcement requirements. However, these reductions are more evenly distributed across the country because, for example, coordinating electric vehicle charging can reduce peak demand anywhere, regardless of climate.

Fig. \ref{barGrapgh} shows nationwide grid reinforcement cost reductions from the three individual strategies and combinations thereof. These estimates use the mean grid reinforcement price of 960 \$/kW, which gives a nationwide cost of \$570 billion (\$180 per household per year) without demand-side management. Individually (top set of bars), the three strategies reduce costs by 19\%, 28\% and 43\%, respectively. Perhaps surprisingly, device coordination -- a strategy based mainly on software, sensing, and communication, rather than equipment upgrades -- reduces costs more than switching to ground-source heat pumps, a significantly more hardware-intensive and expensive strategy. Cost savings from combined strategies (bottom set of bars) are not strictly additive. For example, the combination of less thermal demand from buildings (19\% individually) and ground-source heat pumps (28\%) reduces costs by 40\%. Combining all three demand-side management strategies reduces nationwide grid reinforcement costs by 71\%, to a total of \$165 billion (\$50 per household per year).

\section*{DISCUSSION}

This paper used bottom-up modeling to estimate the physical and economic impacts of electrification of all residential appliances and personal vehicles on distribution grids in each county of the lower 48 United States. The modeling tools are open-source; the cleaned input data and simulation results are free and public. This paper found that distribution grid reinforcement requirements depend mainly on space heating demand, which can increase peak aggregate demand by a factor of four or more in the coldest counties. Without demand-side management, distribution grid reinforcement could cost \$350 to \$790 billion nationwide, or \$2,800 to \$6,400 per household, but smart electrification could eliminate up to two-thirds of these costs.

\subsection*{Sensitivity to modeling choices}

\begin{figure}
\centering
\includegraphics[clip,width=\textwidth]{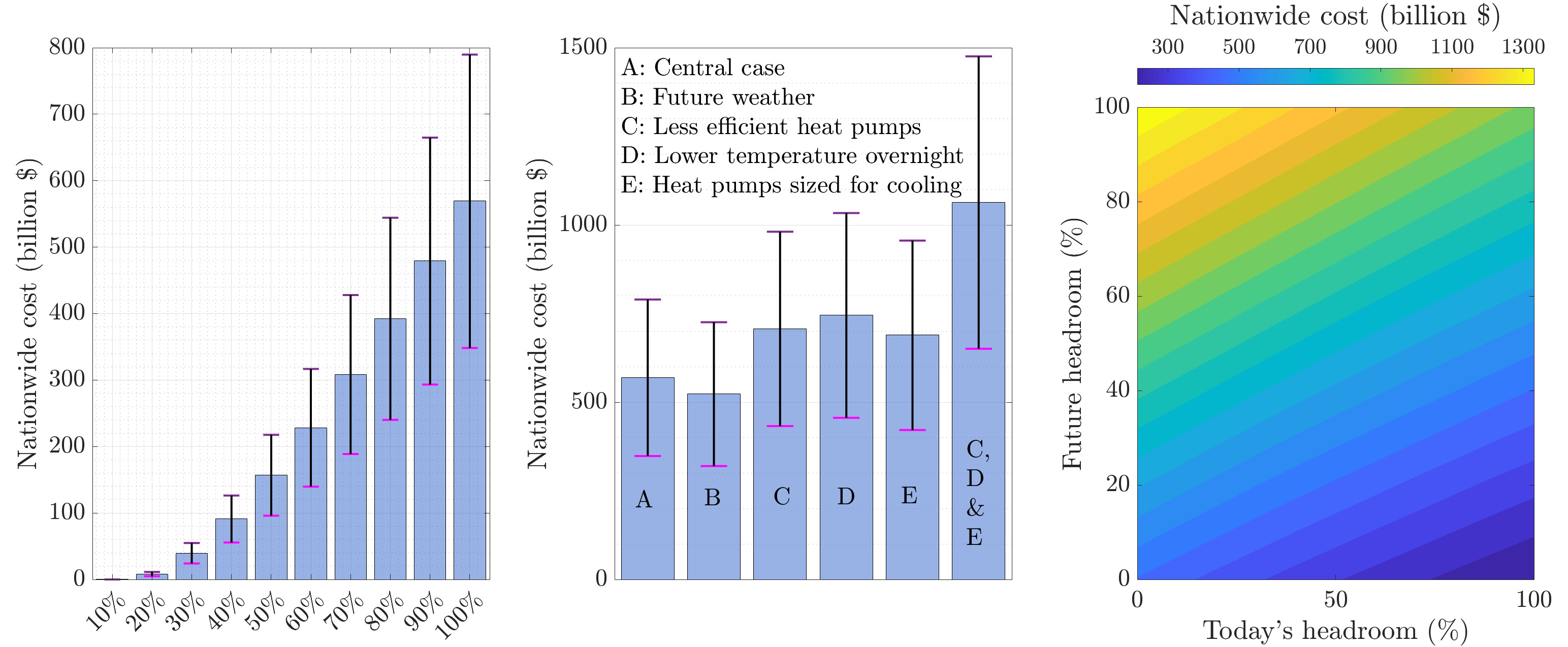}
\caption{Left: Nationwide grid reinforcement costs vs. adoption rate of electric vehicles and heat pumps. Center: Nationwide costs under different modeling choices for space heating. At left and center, bar heights are costs at the mean grid reinforcement price; error bars span costs at the 2.5th to 97.5th price percentiles. Right: Nationwide costs at the mean reinforcement price vs. grid headroom today and in an all-electric future.}\label{combined}
\end{figure}

This section discusses the sensitivity of nationwide distribution grid reinforcement requirements to several modeling assumptions, including heat pump efficiencies, whether households lower their heating temperature setpoints overnight, heat pump sizing protocols, the choice of weather data, the assumed distribution grid headroom, and the electrification adoption rate. Fig. \ref{combined} summarizes the results. The Supplemental Notes discuss the sensitivity analyses in more detail.

Heat pump efficiencies significantly influence grid reinforcement requirements. In cold weather, heat pumps become less efficient and may require electric resistance backup to meet heat demand. The central cases presented above assumed that heat pumps in cold regions met the United States Department of Energy's Cold-Climate Heat Pump Challenge specifications \cite{doe2024cold}. However, in a sensitivity study where heat pumps in cold regions have efficiencies that are similar to today's equipment, nationwide distribution grid reinforcement costs increase by 25\%.

Lowering space heating temperature setpoints overnight can increase grid reinforcement requirements. While lowering setpoints can save energy overnight, warming the thermal mass of a building back up in the morning can require high power and potentially trigger the use of resistance backup heat. In the central cases presented above, users did not alter setpoints overnight. However, in a sensitivity study where users lower setpoints overnight, with randomized but typical setpoint adjustment magnitudes and timings, nationwide distribution grid reinforcement costs increase by 31\%.

Heat pump sizing also influences grid reinforcement requirements, particularly in cold climates. In the central cases presented above, we sized heat pumps to meet the larger of the design heating load and the design cooling load, capping the size at the maximum capacity currently available on the market. If heating demand in very cold weather exceeded the heat pump's capacity, electric resistance backup heat turned on to maintain space temperatures. An alternative approach would be to size heat pumps in colder climates to meet only the design cooling load, relying much more heavily on backup heat in cold weather. We simulated this alternative approach in a sensitivity study and found that it increases nationwide grid reinforcement requirements by 21\%. In the coldest regions (climatic zones 5, 6 and 7), however, sizing heat pumps to the design cooling loads increases grid reinforcement costs by 33\%.

The sensitivities of grid reinforcement requirements to modeling choices on heat pump efficiency (25\%), overnight temperature setpoint adjustments (31\%), and heat pump sizing (21\%) are more than additive. In a joint sensitivity study including all three of the alternative modeling choices discussed above, nationwide grid reinforcement costs by 87\%.

The choice of weather data for the all-electric simulations has relatively little influence on grid reinforcement requirements. In the central cases presented above, we used historical weather data from 2021. An alternative approach would be to use hypothetical future weather data that incorporates the effects of climate change. Climate scientists affiliated with the Intergovernmental Panel on Climate Change have developed frameworks known as the Shared Socioeconomic Pathways (SSP) and Representative Concentration Pathways (RCP) \cite{o2020achievements}. These frameworks provide both quantitative and qualitative descriptions of various societal and environmental development pathways through 2100. For example, in the middle-of-the-road scenario SSP2-RCP4.5\cite{chowdhury_2024_10719179}, global mean surface temperature rises about 2 $^\circ$C by 2050. In a sensitivity study with 2050 SSP2-RCP4.5 weather data, nationwide grid reinforcement costs decrease by 8\%. While climate change increases electricity demand peaks from air conditioning in hot regions, climate change also decreases peaks from space heating in cold regions. In the sensitivity study, the latter effect outweighs the former.

Distribution grid headroom assumptions for the business-as-usual and all-electric scenarios influence grid reinforcement requirement estimates. The central cases presented above used a business-as-usual headroom range of 15\% to 36\% (randomized over counties) and an all-electric future headroom of 20\% for all counties. We ran sensitivity studies, sweeping the headroom in the business-as-usual and all-electric scenarios from 0\% to 100\%. In the extreme case of 100\% headroom in the business-as-usual scenario and 0\% in the all-electric scenario -- meaning today's distribution grids are oversized by a factor of two and all-electric future grids have no safety margins --  nationwide grid reinforcement costs \$215 billion.  At the other extreme of 0\% business-as-usual headroom and 100\% all-electric headroom, nationwide grid reinforcement costs \$1.4 trillion.

The adoption rate of electric vehicles and heat pumps also influences grid reinforcement requirements. In the central case presented above, we estimated that 100\% adoption of electric vehicles and heat pumps would cost  \$570 billion nationwide. We ran sensitivity studies, varying the adoption rate from 0\% to 100\%. Nationwide grid reinforcement costs less than \$1 billion at 10\% adoption and a modest \$40 billion at 30\% adoption. Above 40\% adoption, grid reinforcement requirements increase approximately linearly.

\subsection*{Potential impact}

The findings of this paper could inform decisions about energy system design and operation. Power grid operators might use the spatially-resolved estimates of electricity demand profiles under all-electric scenarios, or the open-source software that generated them, in their capacity expansion planning. Policymakers or utility program administrators might use the demand-side management results to shape electrification incentives. For example, the perhaps surprising effectiveness of device coordination in mitigating grid impacts might warrant making electrification incentives contingent on installing equipment with the sensing, communication, and computing capabilities required for device coordination. Utility regulators might refer to the physical or economic estimates in this paper when evaluating utility requests for rate increases to cover grid reinforcement costs driven by electrification. Finally, the main takeaways of this paper might influence national conversations about the future of United States energy systems.

\subsection*{Limitations and future work}

One challenge in this study was finding reliable, spatially-resolved data on distribution grid topologies and reinforcement prices. Due to a lack of topological data, we did not analyze the impacts of electrification on power quality, such as voltage and frequency regulation. We also modeled aggregate grid reinforcement over wide geographical areas; we did not capture the component-level details of utilities' actual grid reinforcement processes, which span scales ranging from a single building's service drop through substation transformers that serve thousands of buildings. These are possible directions for future work. Due to a lack of spatially-resolved grid reinforcement price data, we used the same price distribution in every United States county in this paper. Higher-resolution price data could enable more precise estimation of distribution grid reinforcement costs. Future work could also assess grid reinforcement costs at the level of utility service territories, rather than counties, to clarify how these costs might affect retail electricity prices.

There are many other opportunities to extend this work. This paper's scope covers distribution grid reinforcement driven by electrification of the residential sector and personal vehicles in the United States. Future research could extend this work to include transmission grid reinforcement and generation capacity expansion; to include electrification of the commercial and industrial sectors; or to include other countries. In particular, estimating transmission and generation costs from residential electrification would be straightforward and useful extensions. Future research could also extend the open-source EDGIE toolbox to incorporate solar photovoltaics, stationary batteries, bidirectional electric vehicle charging, or thermal energy storage. Finally, while this paper estimated the value of demand-side management strategies in mitigating distribution grid reinforcement costs, these strategies could also reduce transmission and generation capacity expansion costs, greenhouse gas emissions, air pollution, and household energy bills. An interesting direction for future work is to identify location-specific mixes of grid upgrades and demand-side management measures that appropriately balance the various costs and benefits. 


\section*{METHODS}
\label{methods}

\subsection*{Data}

Table \ref{tab:dataSources} summarizes the data sources drawn used in this paper. Distributions of building and appliance properties follow the ResStock database \cite{NREL2024BuildingTypology}. Heat pump coefficient of performance curves come from manufacturer data aggregated by the Northeast Energy Efficiency Partnerships \cite{NEEP2023ASHP}. We generate domestic hot water time series from \cite{hpwh}. The number of vehicles per household in each county comes from census data \cite{USCensus2023Vehicles}. We classify personal vehicles as either small (e.g., coupes and sedans) or large (e.g., pickup trucks and sport utility vehicles) and tune the distribution of small and large vehicles in each county to survey data from \cite{iSeecars}. We tune the driving efficiencies of small and large electric vehicles to the Tesla Model S and Ford F-150 lightning, respectively. For both vehicle types, driving efficiencies vary with outdoor temperatures according to \cite{yuksel2015effects}; this variation reflects energy use for cabin heating and cooling, as well as temperature-dependent battery chemistry. Commute distances follow census data for each county \cite{USCensus2023Commuting}. The Supplemental Notes contain further details on modeling, simulation, and data sources. 

\begin{table}[h!]
    \centering
    \caption{Data Sources}
    \begin{tabular}{l | l | l }
        \textbf{Parameter}  &  \textbf{Data Source}  & \textbf{Ref.}  \\
        \hline
        Building types and floor areas &  NREL's ResStock & \cite{reyna2022us, Resstock} \\
        
        Business-as-usual appliances & NREL's ResStock & \cite{reyna2022us, Resstock}\\

        Building insulation and air sealing &  ASHRAE code & \cite{ashrae_90_2_2001_addendum} \\
        
        Design heating and cooling temperatures &  EnergyStar & \cite{EnergyStar} \\
        
        Heat pump coefficients of performance & Northeast Energy Efficiency Partnerships & \cite{NEEP2023ASHP}\\
        
        Domestic hot water profiles & Building America analysis spreadsheets & \cite{hpwh}\\
                
        Commute times &  Census survey& \cite{Census}\\
        
        Vehicles per household &  Census survey & \cite{Census}\\
        
        Electric vehicle parameters & Manufacturer specification and journals & \cite{yuksel2015effects, EV}  \\
        
        Weather data & Oikolab & \cite{oiko}\\
        
        Climate zones & ASHRAE/IECC & \cite{ASHRAE2023Meteo}\\
        
        Grid upgrade prices & Journals  & \cite{rauschkolb2021estimating}\\
        
    \end{tabular}
    \label{tab:dataSources}
\end{table}

\subsection*{Space heating and cooling} 

EDGIE models indoor temperature dynamics using first-order linear ordinary differential equations:
\begin{equation}
\begin{aligned}
     &C_1 \dot T_1(t) =\frac{\theta_1(t) - T_1(t)}{R_1} + q_1(t) + w_1(t) \\
     &0 \leq q_1(t) \leq \overline q_1(t) + \overline p_{1r}.
\end{aligned} \label{buildings}
\end{equation}
Here $t$ (h) denotes time, $T_1$ ($^{\circ}$C) is the indoor air temperature, $C_1$ (kWh/$^{\circ}$C) is the indoor air's thermal capacitance, $\theta_1$ ($^{\circ}$C) is the outdoor air temperature, $R_1$ ($^{\circ}$C/kW) is the thermal resistance between indoor and outdoor air, $q_1$ (kW) is the thermal power supplied by the heat pump and/or resistance heater, $w_1$ (kW) is the exogenous thermal power from the sun, plug loads, lights, body heat, {\it etc.}, $\overline q_1$ (kW) is the heat pump's thermal power capacity, and $\overline p_{1r}$ (kW) is the electric power capacity of the resistance heater. In discrete time, 
\begin{equation}
\begin{aligned}
     T_1(k+1) &= a_1 T_1(k) +(1 - a_1) [\theta_1(k) + R_1(q_1(k) + w_1(k))] ,
\end{aligned}
\end{equation}
where $k$ indexes time steps, $a_1 = \exp(- \Delta t / (R_1 C_1) )$, and $\Delta t$ (h) is the time step duration. Backup resistance heat runs only when the heat pump cannot meet heating demand. The heat pump's control system tries to perfectly track its setpoint $\hat T$ ($^\circ$C) by delivering thermal power
\begin{equation}
\hat q_1(k) = \frac{1}{R_1} \left( \frac{\hat T_1(k+1) - a_1 T_1(k)}{1 - a_1} - \theta_1(k) \right) - w_1(k) ,
\end{equation}
but may saturate at an upper or lower capacity limit. In this model, the total electric power $p_1$ (kW) used by the heat pump and resistance backup is

\begin{equation}
p_1(k) =
\begin{cases} 
0 & \hat{q}_1(k) \leq 0 \\
\hat{q}_1(k) / \eta_1(k)  & 0 < \hat{q}_1(k) \leq \overline q_1(k) \\
\overline q_1(k) / \eta_1(k) + \hat{q}_1(k) - \overline q_1(k) & \overline q_1(k) < \hat{q}_1(k) \leq \overline q_1(k) + \overline p_{1r} \\
\overline q_1(k) / \eta_1(k) + \overline p_{1r} & \overline q_1(k) + \overline p_{1r} < \hat{q}_1(k) ,
\end{cases}
\label{piecewisePower}
\end{equation}
where $\eta_1$ is the heat pump's coefficient of performance. Experimental validation of the space heating model can be found in \cite{priyadarshan2024edgie}. Space cooling works similarly but without resistance backup.

\subsection*{Electric vehicles} 

EDGIE models electric vehicle batteries via
\begin{equation}
\begin{aligned}
     &\dot E(t) = -rE(t) + \eta_2p_2(t)-w_2(t) \\
     &0 \leq E(t) \leq \overline E \\
     &0 \leq p_2(t) \leq \overline p_2.
\end{aligned}
\label{Emax}
\end{equation}
Here $E$ (kWh) is the chemical energy stored in the battery, $r$ (1/h) is the battery's self-dissipation rate, $\eta_1$ is the charging efficiency, $p_1$ (kW) is the electric charging power, $w_1$ (kW) is the chemical power discharged to drive the vehicle, $\overline E$ (kWh) is the energy capacity, and $\overline p_1$ (kW) is the charging power capacity. In discrete time,
\begin{equation}
E(k+1) = a_2 E(k) + \frac{1 - a_2}{r}(\eta_2 p_2(k)-w_2(k)),
\end{equation}
where $a_2 = \exp(-r \Delta t)$.

\subsection*{Water heating} 

EDGIE models water heaters via
\begin{equation}
\begin{aligned}
     &C_3 \dot T_3(t) =\frac{\theta_3 - T_3(t)}{R_3} + q_3(t) - w_3(t) \\
     &0 \leq q_3(t) \leq \eta_3 \overline p_{3h}+ \overline p_{3r}.
\end{aligned} \label{waterHeater}
\end{equation}
Here $T_3$ ($^{\circ}$C) is the water temperature, $C_3$ (kWh/$^{\circ}$C) is
the water's thermal capacitance, $\theta_3$ ($^{\circ}$C) is the (constant) air temperature surrounding the tank, $R_3$ ($^{\circ}$C/kW)
is the thermal resistance between the water and surrounding air, $q_3$ (kW) is the thermal power supplied to the tank, $w_3$ (kW) is the thermal power withdrawn for showers, dish-washing, laundry, {\it etc.}, $\eta_3$ is the water heater's coefficient of performance ($\eta_3 = 1$ for resistance water heaters), $\overline p_{3h}$ (kW) is the heat pump's electric power capacity ($\overline p_{3h} = 0$ for resistance water heaters), and $\overline p_{3r}$ (kW) is the electric power capacity of the resistance heater. In discrete time, 
\begin{equation}
     T_3(k+1) = a_3 T_3(k) +(1 - a_3)[\theta_3 + R_3(q_3(k) - w_3(k))] ,
\end{equation}
where $a_3 = \exp(- \Delta t / (R_3 C_3) )$. The water heating control logic, including dispatch of backup resistance heat for hybrid heat-pump water heaters, is the same as the space heating control logic.

\subsection*{Distribution grid reinforcement costs}

\subsubsection*{Total cost calculation}

\begin{figure}[h]
    \centering
	\includegraphics[ clip,width = 0.7\linewidth]{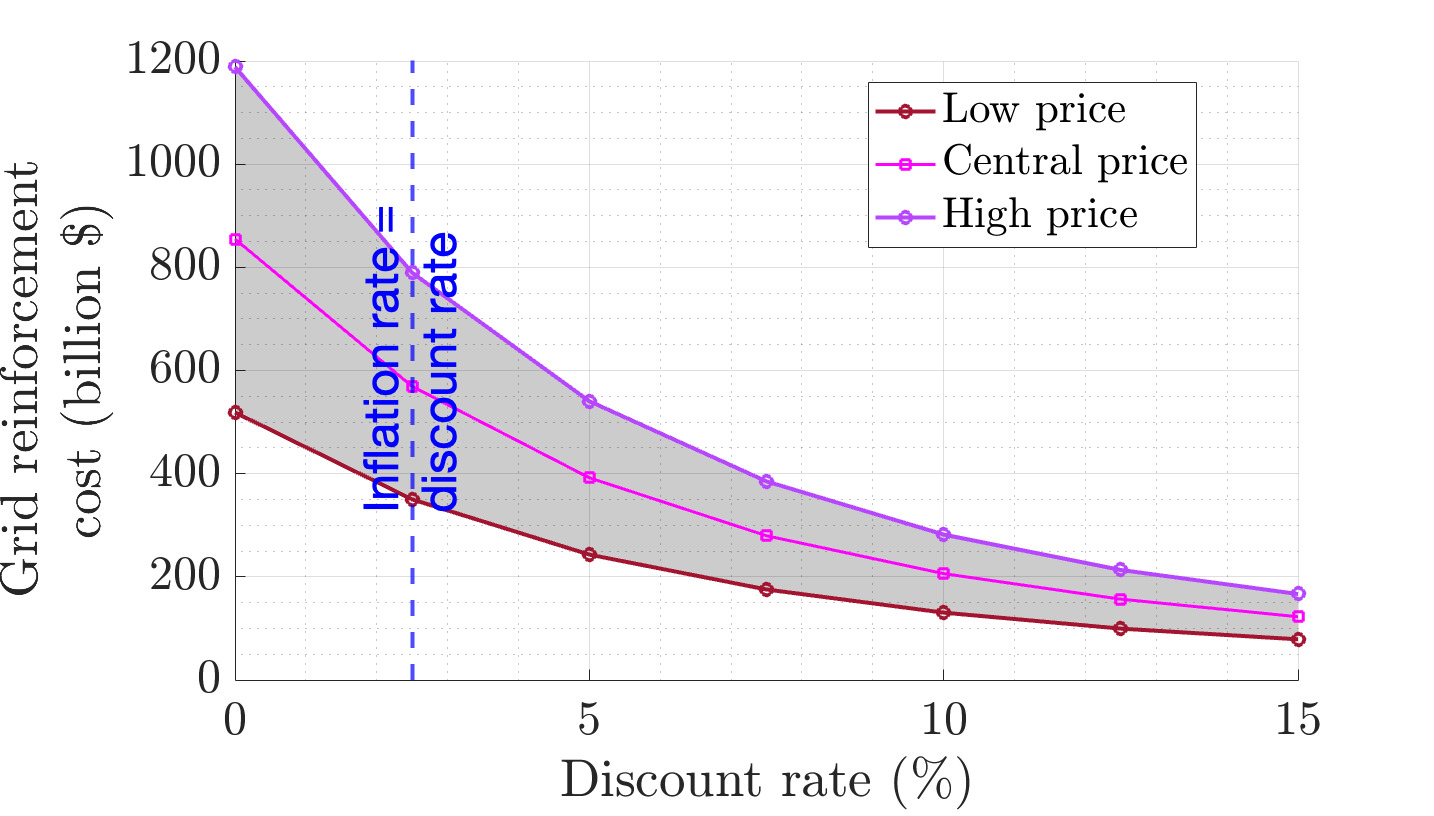}
	\caption{Distribution grid reinforcement cost vs. discount rate, assuming an annual inflation rate of 2.5\%. The shaded gray area is the 95\% confidence region.}
	\label{inflation}
\end{figure}

We assume that distribution grid capacities increase linearly from the business-as-usual capacities in 2025 to the all-electric capacities in 2050. Distribution grid reinforcement costs include both capital costs associated with installing new infrastructure and operation and maintenance cost increases associated with running a larger grid. The net present cost of grid reinforcement is
\begin{equation}
\underbrace{\frac{ \pi_{\text{recurring}} G }{ n } \sum_{k=1}^{n} \frac{ k (1 + i)^k }{ (1 + r)^k}}_{\text{Present value of recurring costs}} 
+ 
\underbrace{\frac{ \pi_{\text{capital}} G }{ n } \sum_{k=1}^{n}\frac{ (1 + i) ^k }{ (1 + r)^k}}_{\text{Present value of capital costs}} .
\label{totalCost}
\end{equation}
Here $\pi_{\text{recurring}}$ (\$/kW) is the initial recurring price in the first year, $\pi_{\text{capital}}$ (\$/kW) is the initial capital price in the first year, $i$ is the annual inflation rate, $r$ is the discount rate, $G$ (kW) is the total growth in grid capacity, $k$ indexes years, $n$ is the total number of years over which recurring costs are projected. Fig. \ref{inflation} shows the distribution grid reinforcement cost vs. the discount rate $r$. To compute the cost estimates reported in the main document, we set the discount rate equal to the annual inflation rate of 2.5\%. Higher discount rates result in lower net present cost estimates.

\subsubsection*{Grid reinforcement requirement uncertainty}
\label{sec:metric}

\begin{figure}
    \centering
	\includegraphics[clip, height=5cm]{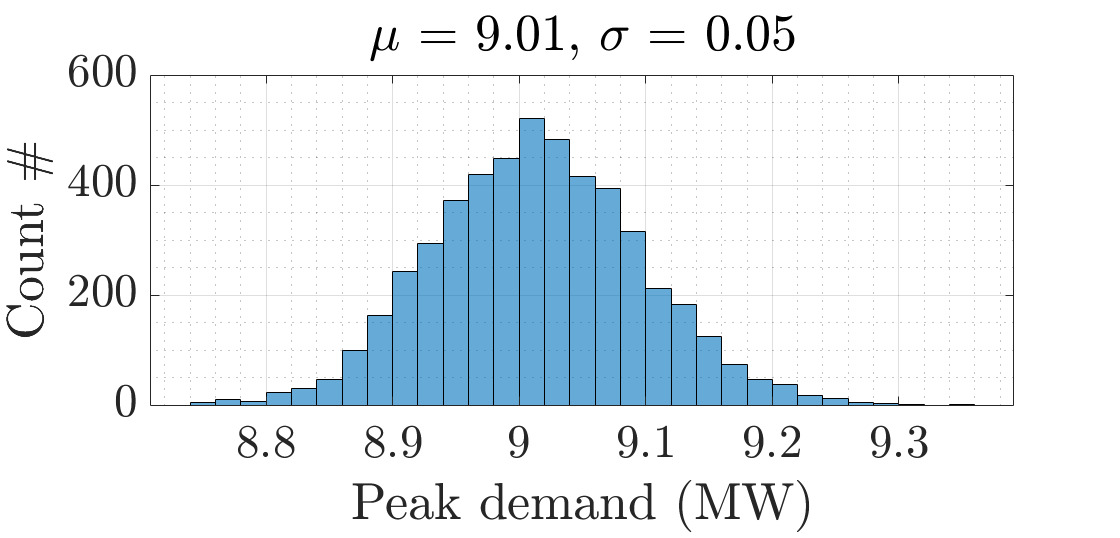}
	\caption{Histogram of the all-electric peak aggregate demand of 1,000 Minnesota homes over 5,000 Monte Carlo simulations. The distribution is approximately Gaussian with a standard deviation of about 0.6\% of the mean.}
	\label{peak}
\end{figure}

In principle, the grid reinforcement requirement $G$ (kW) is a random variable that depends on the randomly generated parameters in the building and machine models. However, $G$ is computed in each county by summing over 1,000 homes and about 2,000 vehicles. From the Central Limit Theorem, therefore, we expect $G$ to have an approximately Gaussian distribution with small variance. Fig. \ref{peak} shows the distribution of the peak aggregate demand for a 1,000-home neighborhood over 5,000 Monte Carlo simulations. The standard deviation of the peak aggregate demand is about 0.6\% of the mean. Given that nationwide simulations require significant computing time and the variations of county-level peaks from one Monte Carlo simulation to the next are small, we conducted the 1,000-home simulations once for each county and treated the resulting grid reinforcement estimate as deterministic.

\subsubsection*{Grid reinforcement cost uncertainty}

We estimate the grid reinforcement cost using the prices $\pi_\text{recurring}$ and $\pi_\text{capital}$ from \cite{rauschkolb2021estimating}. Since \cite{rauschkolb2021estimating} does not model the uncertainty associated with these prices, we assume they follow Gaussian distributions with standard deviations of 20\% of the mean prices. As the total cost in Eq. \eqref{totalCost} is a linear combination of $\pi_\text{recurring}$ and $\pi_\text{capital}$, the total cost is also Gaussian. We calculate its mean and variance analytically using standard formulas.

\section*{RESOURCE AVAILABILITY}


\subsubsection*{Lead contact}


Requests for further information and resources should be directed to and will be fulfilled by the lead contact, Kevin J. Kircher (kircher@purdue.edu).

\subsubsection*{Materials availability}

No materials were used in this study.

\subsubsection*{Data and code availability}
The data and code that support this study are available at 
 https://github.com/priyada7/EDGIE and at https://zenodo.org/records/15857150.

\section*{ACKNOWLEDGMENTS}

This work was supported through the Purdue University Center for High Performance Buildings grant \#60. The authors thank James Braun, Roy Crawford, Andrew Hjortland, Scott Munns, Wayne Craft, Ronald Domitrovic, Elias Pergantis, and Sophia Evers for helpful discussion. The authors thank the anonymous peer reviewers for improving the manuscript through their detailed and insightful feedback.



\section*{SUPPLEMENTAL INFORMATION INDEX}

Document S1. Supplemental experimental procedures, Figures S1–S13, Table S1.

\section*{AUTHOR CONTRIBUTIONS}


Conceptualization: P. and K.J.K; methodology: P., C.C., K.B. and K.J.K; investigation: P.; writing -- original draft: P. and K.J.K; writing -- review \& editing, P., C.C., K.B. and K.J.K; funding acquisition: K.J.K; resources: K.J.K.; supervision: C.C., K.B. and K.J.K.

\section*{DECLARATION OF INTERESTS}

The authors declare no competing interests.

\bibliography{references}

\begin{thebibliography}{62}
\providecommand{\natexlab}[1]{#1}
\providecommand{\url}[1]{\texttt{#1}}
\providecommand{\href}[2]{#2}
\providecommand{\path}[1]{#1}
\providecommand{\DOIprefix}{doi:}
\providecommand{\ArXivprefix}{arXiv:}
\providecommand{\URLprefix}{}
\providecommand{\Pubmedprefix}{pmid:}
\providecommand{\doi}[1]{\href{http://dx.doi.org/#1}{\path{#1}}}
\providecommand{\Pubmed}[1]{\href{pmid:#1}{\path{#1}}}
\providecommand{\BIBand}{and}
\providecommand{\bibinfo}[2]{#2}
\ifx\xfnm\undefined \def\xfnm[#1]{\unskip,\space#1}\fi
\makeatletter\def\@biblabel#1{#1.}\makeatother
\bibitem[{{U.S. Environmental Protection Agency}(2023)}]{USEPA2023}
\bibinfo{author}{{U.S. Environmental Protection Agency}}.
\newblock \bibinfo{title}{Sources of greenhouse gas emissions}.
\newblock
  \bibinfo{howpublished}{\url{https://www.epa.gov/ghgemissions/sources-greenhouse-gas-emissions}}
  (\bibinfo{year}{2023}).
\newblock \bibinfo{note}{Accessed: 2023-04-28}.
\bibitem[{Wilson et~al.(2024)Wilson, Munankarmi, Less, Reyna and
  Rothgeb}]{wilson2024heat}
\bibinfo{author}{Wilson, E.~J.}, \bibinfo{author}{Munankarmi, P.},
  \bibinfo{author}{Less, B.~D.}, \bibinfo{author}{Reyna, J.~L.}, and
  \bibinfo{author}{Rothgeb, S.} (\bibinfo{year}{2024}). \bibinfo{title}{Heat
  pumps for all? {Distributions} of the costs and benefits of residential
  air-source heat pumps in the {{United States}}}.
\newblock \bibinfo{journal}{Joule}.
\bibitem[{{{California} Air Resources Board}(2017)}]{carb_2017}
\bibinfo{author}{{{California} Air Resources Board}}.
\newblock \bibinfo{title}{{California}'s 2017 climate change scoping plan}.
\newblock
  \bibinfo{howpublished}{\url{https://ww2.arb.ca.gov/our-work/programs/ab-32-climate-change-scoping-plan/2017-scoping-plan-documents}}
  (\bibinfo{year}{2017}).
\newblock \bibinfo{note}{Accessed: 2024-08-29}.
\bibitem[{{New York State Energy Research and Development
  Authority}(2024)}]{nyserda_heat_pumps_2024}
\bibinfo{author}{{New York State Energy Research and Development Authority}}.
\newblock \bibinfo{title}{Heat pumps outsell gas furnaces again}.
\newblock
  \bibinfo{howpublished}{\url{https://www.nyserda.ny.gov/Featured-Stories/Heat-Pumps-Outsell-Gas-Furnaces-Again}}
  (\bibinfo{year}{2024}).
\newblock \bibinfo{note}{Accessed: 2024-08-29}.
\bibitem[{Keskar et~al.(2023)Keskar, Galik and Johnson}]{keskar2023planning}
\bibinfo{author}{Keskar, A.}, \bibinfo{author}{Galik, C.}, and
  \bibinfo{author}{Johnson, J.~X.} (\bibinfo{year}{2023}).
  \bibinfo{title}{Planning for winter peaking power systems in the {{United
  States}}}.
\newblock \bibinfo{journal}{Energy Policy} \emph{\bibinfo{volume}{173}},
  \bibinfo{pages}{113376}.
\bibitem[{Mai et~al.(2018)Mai, Jadun, Logan, McMillan, Muratori, Steinberg,
  Vimmerstedt, Haley, Jones and Nelson}]{mai2018electrification}
\bibinfo{author}{Mai, T.~T.}, \bibinfo{author}{Jadun, P.},
  \bibinfo{author}{Logan, J.~S.}, \bibinfo{author}{McMillan, C.~A.},
  \bibinfo{author}{Muratori, M.}, \bibinfo{author}{Steinberg, D.~C.},
  \bibinfo{author}{Vimmerstedt, L.~J.}, \bibinfo{author}{Haley, B.},
  \bibinfo{author}{Jones, R.}, and \bibinfo{author}{Nelson, B.}
\newblock \bibinfo{title}{Electrification futures study: Scenarios of electric
  technology adoption and power consumption for the {{United States}}}.
\newblock \bibinfo{type}{Tech. Rep.} National Renewable Energy Lab.(NREL),
  Golden, CO ({{United States}}) (\bibinfo{year}{2018}).
\bibitem[{Tarroja et~al.(2018)Tarroja, Chiang, AghaKouchak, Samuelsen,
  Raghavan, Wei, Sun and Hong}]{tarroja2018translating}
\bibinfo{author}{Tarroja, B.}, \bibinfo{author}{Chiang, F.},
  \bibinfo{author}{AghaKouchak, A.}, \bibinfo{author}{Samuelsen, S.},
  \bibinfo{author}{Raghavan, S.~V.}, \bibinfo{author}{Wei, M.},
  \bibinfo{author}{Sun, K.}, and \bibinfo{author}{Hong, T.}
  (\bibinfo{year}{2018}). \bibinfo{title}{Translating climate change and
  heating system electrification impacts on building energy use to future
  greenhouse gas emissions and electric grid capacity requirements in
  {C}alifornia}.
\newblock \bibinfo{journal}{Applied Energy} \emph{\bibinfo{volume}{225}},
  \bibinfo{pages}{522--534}.
\bibitem[{Amonkar et~al.(2023)Amonkar, Doss-Gollin, Farnham, Modi and
  Lall}]{amonkar2023differential}
\bibinfo{author}{Amonkar, Y.}, \bibinfo{author}{Doss-Gollin, J.},
  \bibinfo{author}{Farnham, D.~J.}, \bibinfo{author}{Modi, V.}, and
  \bibinfo{author}{Lall, U.} (\bibinfo{year}{2023}).
  \bibinfo{title}{Differential effects of climate change on average and peak
  demand for heating and cooling across the contiguous {{USA}}}.
\newblock \bibinfo{journal}{Communications Earth \& Environment}
  \emph{\bibinfo{volume}{4}}, \bibinfo{pages}{402}.
\bibitem[{Buonocore et~al.(2022)Buonocore, Salimifard, Magavi and
  Allen}]{buonocore2022inefficient}
\bibinfo{author}{Buonocore, J.~J.}, \bibinfo{author}{Salimifard, P.},
  \bibinfo{author}{Magavi, Z.}, and \bibinfo{author}{Allen, J.~G.}
  (\bibinfo{year}{2022}). \bibinfo{title}{Inefficient building electrification
  will require massive buildout of renewable energy and seasonal energy
  storage}.
\newblock \bibinfo{journal}{Scientific Reports} \emph{\bibinfo{volume}{12}},
  \bibinfo{pages}{11931}.
\bibitem[{Bistline et~al.(2021)Bistline, Roney, McCollum and
  Blanford}]{bistline2021deep}
\bibinfo{author}{Bistline, J.~E.}, \bibinfo{author}{Roney, C.~W.},
  \bibinfo{author}{McCollum, D.~L.}, and \bibinfo{author}{Blanford, G.~J.}
  (\bibinfo{year}{2021}). \bibinfo{title}{Deep decarbonization impacts on
  electric load shapes and peak demand}.
\newblock \bibinfo{journal}{Environmental Research Letters}
  \emph{\bibinfo{volume}{16}}, \bibinfo{pages}{094054}.
\bibitem[{Ssembatya et~al.(2024)Ssembatya, Kern, Oikonomou, Voisin, Burleyson
  and Akdemir}]{ssembatya2024dual}
\bibinfo{author}{Ssembatya, H.}, \bibinfo{author}{Kern, J.~D.},
  \bibinfo{author}{Oikonomou, K.}, \bibinfo{author}{Voisin, N.},
  \bibinfo{author}{Burleyson, C.~D.}, and \bibinfo{author}{Akdemir, K.~Z.}
  (\bibinfo{year}{2024}). \bibinfo{title}{Dual impacts of space heating
  electrification and climate change increase uncertainties in peak load
  behavior and grid capacity requirements in {Texas}}.
\newblock \bibinfo{journal}{Earth's Future} \emph{\bibinfo{volume}{12}},
  \bibinfo{pages}{e2024EF004443}.
\bibitem[{Protopapadaki and Saelens(2017)}]{protopapadaki2017heat}
\bibinfo{author}{Protopapadaki, C.}, and \bibinfo{author}{Saelens, D.}
  (\bibinfo{year}{2017}). \bibinfo{title}{Heat pump and {PV} impact on
  residential low-voltage distribution grids as a function of building and
  district properties}.
\newblock \bibinfo{journal}{Applied Energy} \emph{\bibinfo{volume}{192}},
  \bibinfo{pages}{268--281}.
\bibitem[{Al-Awami et~al.(2016)Al-Awami, Sortomme, Asim~Akhtar and
  Faddel}]{7275178}
\bibinfo{author}{Al-Awami, A.~T.}, \bibinfo{author}{Sortomme, E.},
  \bibinfo{author}{Asim~Akhtar, G.~M.}, and \bibinfo{author}{Faddel, S.}
  (\bibinfo{year}{2016}). \bibinfo{title}{A voltage-based controller for an
  electric-vehicle charger}.
\newblock \bibinfo{journal}{IEEE Transactions on Vehicular Technology}
  \emph{\bibinfo{volume}{65}}, \bibinfo{pages}{4185--4196}.
  \DOIprefix\doi{10.1109/TVT.2015.2481712}.
\bibitem[{Damianakis et~al.(2023)Damianakis, Mouli, Bauer and
  Yu}]{damianakis2023assessing}
\bibinfo{author}{Damianakis, N.}, \bibinfo{author}{Mouli, G. R.~C.},
  \bibinfo{author}{Bauer, P.}, and \bibinfo{author}{Yu, Y.}
  (\bibinfo{year}{2023}). \bibinfo{title}{Assessing the grid impact of electric
  vehicles, heat pumps \& {PV} generation in {D}utch {LV} distribution grids}.
\newblock \bibinfo{journal}{Applied Energy} \emph{\bibinfo{volume}{352}},
  \bibinfo{pages}{121878}.
\bibitem[{McKenna et~al.(2024)McKenna, Abraham and Wang}]{mckenna2024major}
\bibinfo{author}{McKenna, K.}, \bibinfo{author}{Abraham, S.~A.}, and
  \bibinfo{author}{Wang, W.}
\newblock \bibinfo{title}{Major drivers of long-term distribution transformer
  demand}.
\newblock \bibinfo{type}{Tech. Rep.} National Renewable Energy Laboratory
  (NREL), Golden, CO ({{United States}}) (\bibinfo{year}{2024}).
\bibitem[{Larsen et~al.(2018)Larsen, Boehlert, Eto, Hamachi-LaCommare,
  Martinich and Rennels}]{larsen2018projecting}
\bibinfo{author}{Larsen, P.~H.}, \bibinfo{author}{Boehlert, B.},
  \bibinfo{author}{Eto, J.}, \bibinfo{author}{Hamachi-LaCommare, K.},
  \bibinfo{author}{Martinich, J.}, and \bibinfo{author}{Rennels, L.}
  (\bibinfo{year}{2018}). \bibinfo{title}{Projecting future costs to {US}
  electric utility customers from power interruptions}.
\newblock \bibinfo{journal}{Energy} \emph{\bibinfo{volume}{147}},
  \bibinfo{pages}{1256--1277}.
\bibitem[{Fant et~al.(2020)Fant, Boehlert, Strzepek, Larsen, White, Gulati, Li
  and Martinich}]{fant2020climate}
\bibinfo{author}{Fant, C.}, \bibinfo{author}{Boehlert, B.},
  \bibinfo{author}{Strzepek, K.}, \bibinfo{author}{Larsen, P.},
  \bibinfo{author}{White, A.}, \bibinfo{author}{Gulati, S.},
  \bibinfo{author}{Li, Y.}, and \bibinfo{author}{Martinich, J.}
  (\bibinfo{year}{2020}). \bibinfo{title}{Climate change impacts and costs to
  us electricity transmission and distribution infrastructure}.
\newblock \bibinfo{journal}{Energy} \emph{\bibinfo{volume}{195}},
  \bibinfo{pages}{116899}.
\bibitem[{{Synapse Energy Economics}(2019)}]{synapseEVsDrivingRatesDown2019}
\bibinfo{author}{{Synapse Energy Economics}}.
\newblock \bibinfo{title}{{EVs} are driving rates down}.
\newblock
  \bibinfo{howpublished}{\url{https://www.synapse-energy.com/evs-are-driving-rates-down}}
  (\bibinfo{year}{2019}).
\bibitem[{Steinberg et~al.(2017)Steinberg, Bielen, Eichman, Eurek, Logan, Mai,
  McMillan, Parker, Vimmerstedt and Wilson}]{steinberg2017electrification}
\bibinfo{author}{Steinberg, D.}, \bibinfo{author}{Bielen, D.},
  \bibinfo{author}{Eichman, J.}, \bibinfo{author}{Eurek, K.},
  \bibinfo{author}{Logan, J.}, \bibinfo{author}{Mai, T.},
  \bibinfo{author}{McMillan, C.}, \bibinfo{author}{Parker, A.},
  \bibinfo{author}{Vimmerstedt, L.}, and \bibinfo{author}{Wilson, E.}
\newblock \bibinfo{title}{Electrification and decarbonization: exploring us
  energy use and greenhouse gas emissions in scenarios with widespread
  electrification and power sector decarbonization}.
\newblock \bibinfo{type}{Tech. Rep.} National Renewable Energy Lab.(NREL),
  Golden, CO ({United States}) (\bibinfo{year}{2017}).
\bibitem[{Murphy et~al.(2021)Murphy, Mai, Sun, Jadun, Muratori, Nelson and
  Jones}]{murphy2021electrification}
\bibinfo{author}{Murphy, C.}, \bibinfo{author}{Mai, T.}, \bibinfo{author}{Sun,
  Y.}, \bibinfo{author}{Jadun, P.}, \bibinfo{author}{Muratori, M.},
  \bibinfo{author}{Nelson, B.}, and \bibinfo{author}{Jones, R.}
\newblock \bibinfo{title}{Electrification futures study: Scenarios of power
  system evolution and infrastructure development for the {United States}}.
\newblock \bibinfo{type}{Tech. Rep.} National Renewable Energy Lab.(NREL),
  Golden, CO ({United States}) (\bibinfo{year}{2021}).
\bibitem[{{Electric Power Research Institute}(2018)}]{epri}
\bibinfo{author}{{Electric Power Research Institute}}.
\newblock \bibinfo{title}{{U.S. National Electrification Assessment}}.
\newblock \bibinfo{type}{Tech. Rep.} (\bibinfo{year}{2018}).
\bibitem[{Reyna et~al.(2022)Reyna, Wilson, Parker, Satre-Meloy, Egerter,
  Bianchi, Praprost, Speake, Liu, Horsey et~al.}]{reyna2022us}
\bibinfo{author}{Reyna, J.}, \bibinfo{author}{Wilson, E.},
  \bibinfo{author}{Parker, A.}, \bibinfo{author}{Satre-Meloy, A.},
  \bibinfo{author}{Egerter, A.}, \bibinfo{author}{Bianchi, C.},
  \bibinfo{author}{Praprost, M.}, \bibinfo{author}{Speake, A.},
  \bibinfo{author}{Liu, L.}, \bibinfo{author}{Horsey, R.} et~al.
\newblock \bibinfo{title}{{US} building stock characterization study: A
  national typology for decarbonizing {US} buildings}.
\newblock \bibinfo{type}{Tech. Rep.} National Renewable Energy Lab.(NREL),
  Golden, CO ({United States}) (\bibinfo{year}{2022}).
\bibitem[{Specian et~al.(2021)Specian, Cohn and York}]{Specian2021}
\bibinfo{author}{Specian, M.}, \bibinfo{author}{Cohn, C.}, and
  \bibinfo{author}{York, D.}
\newblock \bibinfo{title}{Demand-side solutions to winter peaks and
  constraints}.
\newblock \bibinfo{type}{Research Report} American Council for an
  Energy-Efficient Economy (ACEEE) \bibinfo{address}{Washington, DC}
  (\bibinfo{year}{2021}).
\newblock \URLprefix \url{www.aceee.org/research-report/u2101}.
\bibitem[{White et~al.(2021)White, Rhodes, Wilson and
  Webber}]{white2021quantifying}
\bibinfo{author}{White, P.~R.}, \bibinfo{author}{Rhodes, J.~D.},
  \bibinfo{author}{Wilson, E.~J.}, and \bibinfo{author}{Webber, M.~E.}
  (\bibinfo{year}{2021}). \bibinfo{title}{Quantifying the impact of residential
  space heating electrification on the {T}exas electric grid}.
\newblock \bibinfo{journal}{Applied Energy} \emph{\bibinfo{volume}{298}},
  \bibinfo{pages}{117113}.
\bibitem[{Zhang et~al.(2020{\natexlab{a}})Zhang, Greenblatt, MacDougall, Saxena
  and Prabhakar}]{zhang2020quantifying}
\bibinfo{author}{Zhang, C.}, \bibinfo{author}{Greenblatt, J.~B.},
  \bibinfo{author}{MacDougall, P.}, \bibinfo{author}{Saxena, S.}, and
  \bibinfo{author}{Prabhakar, A.~J.} (\bibinfo{year}{2020}{\natexlab{a}}).
  \bibinfo{title}{Quantifying the benefits of electric vehicles on the future
  electricity grid in the midwestern {U}nited {S}tates}.
\newblock \bibinfo{journal}{Applied Energy} \emph{\bibinfo{volume}{270}},
  \bibinfo{pages}{115174}.
\bibitem[{Elmallah et~al.(2022)Elmallah, Brockway and
  Callaway}]{elmallah2022can}
\bibinfo{author}{Elmallah, S.}, \bibinfo{author}{Brockway, A.~M.}, and
  \bibinfo{author}{Callaway, D.} (\bibinfo{year}{2022}). \bibinfo{title}{Can
  distribution grid infrastructure accommodate residential electrification and
  electric vehicle adoption in {Northern {California}?}}
\newblock \bibinfo{journal}{Environmental Research: Infrastructure and
  Sustainability} \emph{\bibinfo{volume}{2}}, \bibinfo{pages}{045005}.
\bibitem[{Li and Jenn(2024)}]{li2024impact}
\bibinfo{author}{Li, Y.}, and \bibinfo{author}{Jenn, A.}
  (\bibinfo{year}{2024}). \bibinfo{title}{Impact of electric vehicle charging
  demand on power distribution grid congestion}.
\newblock \bibinfo{journal}{Proceedings of the National Academy of Sciences}
  \emph{\bibinfo{volume}{121}}, \bibinfo{pages}{e2317599121}.
\bibitem[{{New York State Energy Research and Development Authority
  (NYSERDA)}(2022)}]{nyserda2022_tedi}
\bibinfo{author}{{New York State Energy Research and Development Authority
  (NYSERDA)}}.
\newblock \bibinfo{title}{Transportation electrification distribution system
  impact study}.
\newblock \bibinfo{type}{Final Report} \bibinfo{number}{Report No.\ 22‑13}
  NYSERDA \bibinfo{address}{Albany, NY \& San Francisco, CA}
  (\bibinfo{year}{2022}).
\newblock \URLprefix
  \url{https://www.nyserda.ny.gov/-/media/Project/Nyserda/Files/Publications/Research/Transportation/22-13-Transportation-Electricification-Distribution-System-Impact-Study.pdf}
  \bibinfo{note}{prepared for NYSERDA by Resource Innovations (formerly Nexant
  Inc.) :contentReference[oaicite:0]{index=0}}.
\bibitem[{{Energy + Environmental Economics (E3) \&
  GridLab}(2021)}]{ethree2021_distribution_costs}
\bibinfo{author}{{Energy + Environmental Economics (E3) \& GridLab}}.
\newblock \bibinfo{title}{Distribution grid cost impacts driven by
  transportation electrification}.
\newblock \bibinfo{type}{Research Report} GridLab / E3 (\bibinfo{year}{2021}).
\newblock \URLprefix
  \url{https://www.ethree.com/wp-content/uploads/2021/06/GridLab_2035-Transportation-Dist-Cost.pdf}.
\bibitem[{Ebrahimi et~al.(2018)Ebrahimi, Mac~Kinnon and
  Brouwer}]{ebrahimi2018California}
\bibinfo{author}{Ebrahimi, S.}, \bibinfo{author}{Mac~Kinnon, M.}, and
  \bibinfo{author}{Brouwer, J.} (\bibinfo{year}{2018}).
  \bibinfo{title}{{California} end-use electrification impacts on carbon
  neutrality and clean air}.
\newblock \bibinfo{journal}{Applied energy} \emph{\bibinfo{volume}{213}},
  \bibinfo{pages}{435--449}.
\bibitem[{Wei et~al.(2013)Wei, Nelson, Greenblatt, Mileva, Johnston, Ting,
  Yang, Jones, McMahon and Kammen}]{wei2013deep}
\bibinfo{author}{Wei, M.}, \bibinfo{author}{Nelson, J.~H.},
  \bibinfo{author}{Greenblatt, J.~B.}, \bibinfo{author}{Mileva, A.},
  \bibinfo{author}{Johnston, J.}, \bibinfo{author}{Ting, M.},
  \bibinfo{author}{Yang, C.}, \bibinfo{author}{Jones, C.},
  \bibinfo{author}{McMahon, J.~E.}, and \bibinfo{author}{Kammen, D.~M.}
  (\bibinfo{year}{2013}). \bibinfo{title}{Deep carbon reductions in
  {California} require electrification and integration across economic
  sectors}.
\newblock \bibinfo{journal}{Environmental Research Letters}
  \emph{\bibinfo{volume}{8}}, \bibinfo{pages}{014038}.
\bibitem[{Priyadarshan et~al.(2024)Priyadarshan, Pergantis, Crozier, Baker and
  Kircher}]{priyadarshan2024edgie}
\bibinfo{author}{Priyadarshan}, \bibinfo{author}{Pergantis, E.~N.},
  \bibinfo{author}{Crozier, C.}, \bibinfo{author}{Baker, K.}, and
  \bibinfo{author}{Kircher, K.~J.} (\bibinfo{year}{2024}).
  \bibinfo{title}{{EDGIE}: A simulation test-bed for investigating the impacts
  of building and vehicle electrification on distribution grids}.
\newblock \bibinfo{journal}{Proceedings of the 57th Hawaii International
  Conference on System Sciences}.
\bibitem[{oik(????)}]{oiko}
\bibinfo{title}{Oikolab: Weather and climate data for analysis}.
\newblock \bibinfo{howpublished}{\url{https://www.oikolab.com/}} ().
\bibitem[{REC(2020)}]{RECS}
\bibinfo{title}{Residential energy consumption survey {(RECS)}}.
\newblock
  \bibinfo{howpublished}{\url{https://www.eia.gov/consumption/residential/data/2020/}}
  (\bibinfo{year}{2020}).
\bibitem[{hpw(????)}]{hpwh}
\bibinfo{title}{{Building America Analysis Spreadsheets}}.
\newblock
  \bibinfo{howpublished}{\url{https://www.energy.gov/eere/buildings/building-america-analysis-spreadsheets}}
  ().
\bibitem[{Meinrenken et~al.(2020)Meinrenken, Rauschkolb, Abrol, Chakrabarty,
  Decalf, Hidey, McKeown, Mehmani, Modi and Culligan}]{meinrenken2020mfred}
\bibinfo{author}{Meinrenken, C.~J.}, \bibinfo{author}{Rauschkolb, N.},
  \bibinfo{author}{Abrol, S.}, \bibinfo{author}{Chakrabarty, T.},
  \bibinfo{author}{Decalf, V.~C.}, \bibinfo{author}{Hidey, C.},
  \bibinfo{author}{McKeown, K.}, \bibinfo{author}{Mehmani, A.},
  \bibinfo{author}{Modi, V.}, and \bibinfo{author}{Culligan, P.~J.}
  (\bibinfo{year}{2020}). \bibinfo{title}{{MFRED}, 10 second interval real and
  reactive power for groups of 390 {US} apartments of varying size and
  vintage}.
\newblock \bibinfo{journal}{Scientific Data} \emph{\bibinfo{volume}{7}},
  \bibinfo{pages}{375}.
\bibitem[{{U.S. Department of Energy}(2024)}]{doe2024cold}
\bibinfo{author}{{U.S. Department of Energy}}.
\newblock \bibinfo{title}{{Residential Cold Climate Heat Pump Technology
  Challenge Fact Sheet}} (\bibinfo{year}{2024}).
\newblock \URLprefix
  \url{https://www.energy.gov/eere/buildings/articles/residential-cold-climate-heat-pump-technology-challenge-fact-sheet}
  \bibinfo{note}{accessed: 2024-09-18}.
\bibitem[{Takahashi et~al.(2024)Takahashi, Hopkins, Carson, Schadler and
  Chavin}]{Takahashi2024}
\bibinfo{author}{Takahashi, K.}, \bibinfo{author}{Hopkins, A.},
  \bibinfo{author}{Carson, E.}, \bibinfo{author}{Schadler, S.}, and
  \bibinfo{author}{Chavin, S.}
\newblock \bibinfo{title}{Assessment of electric grid headroom for
  accommodating building electrification}.
\newblock \bibinfo{type}{Technical Report} Synapse Energy Economics, Inc.
  (\bibinfo{year}{2024}).
\bibitem[{Zhang et~al.(2020{\natexlab{b}})Zhang, Jenkins and
  Larson}]{zhang2020princeton}
\bibinfo{author}{Zhang, C.}, \bibinfo{author}{Jenkins, J.}, and
  \bibinfo{author}{Larson, E.~D.}
\newblock \bibinfo{title}{Princeton’s net-zero {America} study annex g:
  Electricity distribution system transition}.
\newblock \bibinfo{type}{Tech. Rep.} Tech. Rep., Princeton University,
  Princeton, NJ (\bibinfo{year}{2020}{\natexlab{b}}).
\bibitem[{Rauschkolb et~al.(2021)Rauschkolb, Limandibhratha, Modi and
  Mercadal}]{rauschkolb2021estimating}
\bibinfo{author}{Rauschkolb, N.}, \bibinfo{author}{Limandibhratha, N.},
  \bibinfo{author}{Modi, V.}, and \bibinfo{author}{Mercadal, I.}
  (\bibinfo{year}{2021}). \bibinfo{title}{Estimating electricity distribution
  costs using historical data}.
\newblock \bibinfo{journal}{Utilities Policy} \emph{\bibinfo{volume}{73}},
  \bibinfo{pages}{101309}.
\bibitem[{{U.S. Energy Information Administration}(2019)}]{EIA_AEO2019}
\bibinfo{author}{{U.S. Energy Information Administration}}.
\newblock \bibinfo{title}{{Annual Energy Outlook 2019}}.
\newblock \bibinfo{howpublished}{{U.S. Department of Energy}}
  (\bibinfo{year}{2019}).
\newblock \URLprefix \url{https://www.eia.gov/outlooks/archive/aeo19/}
  \bibinfo{note}{accessed: 24 January 2019}.
\bibitem[{Fares and King(2017)}]{oedi_489}
\bibinfo{author}{Fares, R.}, and \bibinfo{author}{King, C.}
\newblock \bibinfo{title}{{FERC} {F}orm 1: {E}lectric utility cost, energy
  sales, peak demand, and customer count data 1994-2019}
  (\bibinfo{year}{2017}).
\newblock \URLprefix \url{https://data.openei.org/submissions/489}.
\bibitem[{{National Academies of Sciences, Engineering, and
  Medicine}(2021)}]{E3_DER_Grid_Impacts}
\bibinfo{author}{{National Academies of Sciences, Engineering, and Medicine}}.
\newblock \bibinfo{title}{Accelerating decarbonization of the {U.S.} energy
  system} (\bibinfo{year}{2021}).
\newblock \URLprefix
  \url{https://www.nationalacademies.org/documents/embed/link/LF2255DA3DD1C41C0A42D3BEF0989ACAECE3053A6A9B/file/D01A9C1B61B4008B83CF424B535759678E8A9BB2BF4B?noSaveAs=1}.
\bibitem[{Pergantis et~al.(2025)Pergantis, Premer, Lee, Priyadarshan, Liu,
  Groll, Ziviani, Kircher et~al.}]{pergantis2025protecting}
\bibinfo{author}{Pergantis, E.~N.}, \bibinfo{author}{Premer, L. D.~R.},
  \bibinfo{author}{Lee, A.~H.}, \bibinfo{author}{Priyadarshan},
  \bibinfo{author}{Liu, H.}, \bibinfo{author}{Groll, E.~A.},
  \bibinfo{author}{Ziviani, D.}, \bibinfo{author}{Kircher, K.~J.} et~al.
  (\bibinfo{year}{2025}). \bibinfo{title}{Protecting residential electrical
  panels and service through model predictive control: A field study}.
\newblock \bibinfo{journal}{Applied Energy} \emph{\bibinfo{volume}{386}},
  \bibinfo{pages}{125528}.
\bibitem[{{{U.S.} Energy Information
  Administration}(2020)}]{eia_table_ce11_2020}
\bibinfo{author}{{{U.S.} Energy Information Administration}}.
\newblock \bibinfo{title}{{Table CE1.1 Summary: Total Energy Consumption,
  Expenditures, and Intensities}}.
\newblock
  \bibinfo{howpublished}{\url{https://www.eia.gov/consumption/residential/data/2020/index.php?view=consumption\#summary}}
  (\bibinfo{year}{2020}).
\newblock \URLprefix
  \url{https://www.eia.gov/consumption/residential/data/2020/index.php?view=consumption\#summary}
  \bibinfo{note}{accessed: 2024-09-18}.
\bibitem[{Temple and Crownhart(2022)}]{technologyreview2022climate}
\bibinfo{author}{Temple, J.}, and \bibinfo{author}{Crownhart, C.}
  (\bibinfo{year}{2022}). \bibinfo{title}{{Technology wins in breakthrough
  climate bill}}.
\newblock \bibinfo{journal}{MIT Technology Review}. \URLprefix
  \url{https://www.technologyreview.com/2022/07/28/1056531/technology-wins-breakthrough-climate-bill/}.
\bibitem[{Defense.gov(2022)}]{dod2022defense}
\bibinfo{author}{Defense.gov} (\bibinfo{year}{2022}). \bibinfo{title}{{DOD
  Supports Climate Adaptation, Resilience Through Action Plan}}.
\newblock \URLprefix
  \url{https://www.defense.gov/News/News-Stories/Article/Article/3252968/}.
\bibitem[{Maxim and Grubert(2023)}]{maxim2023highly}
\bibinfo{author}{Maxim, A.}, and \bibinfo{author}{Grubert, E.}
  (\bibinfo{year}{2023}). \bibinfo{title}{Highly energy efficient housing can
  reduce peak load and increase safety under beneficial electrification}.
\newblock \bibinfo{journal}{Environmental Research Letters}
  \emph{\bibinfo{volume}{19}}, \bibinfo{pages}{014036}.
\bibitem[{O’Neill et~al.(2020)O’Neill, Carter, Ebi, Harrison,
  Kemp-Benedict, Kok, Kriegler, Preston, Riahi, Sillmann
  et~al.}]{o2020achievements}
\bibinfo{author}{O’Neill, B.~C.}, \bibinfo{author}{Carter, T.~R.},
  \bibinfo{author}{Ebi, K.}, \bibinfo{author}{Harrison, P.~A.},
  \bibinfo{author}{Kemp-Benedict, E.}, \bibinfo{author}{Kok, K.},
  \bibinfo{author}{Kriegler, E.}, \bibinfo{author}{Preston, B.~L.},
  \bibinfo{author}{Riahi, K.}, \bibinfo{author}{Sillmann, J.} et~al.
  (\bibinfo{year}{2020}). \bibinfo{title}{Achievements and needs for the
  climate change scenario framework}.
\newblock \bibinfo{journal}{Nature Climate Change} \emph{\bibinfo{volume}{10}},
  \bibinfo{pages}{1074--1084}.
\bibitem[{Chowdhury et~al.(2024)Chowdhury, Li, Stubbings, New, Rastogi and
  Kao}]{chowdhury_2024_10719179}
\bibinfo{author}{Chowdhury, S.}, \bibinfo{author}{Li, F.},
  \bibinfo{author}{Stubbings, A.}, \bibinfo{author}{New, J.},
  \bibinfo{author}{Rastogi, D.}, and \bibinfo{author}{Kao, S.-C.}
\newblock \bibinfo{title}{{Future Typical Meteorological Year (fTMY) US Weather
  Files for Building Simulation for every US County in CONUS (Cross-Model
  Version-SSP2-RCP4.5)}} (\bibinfo{year}{2024}).
\newblock \URLprefix \url{https://doi.org/10.5281/zenodo.10719179}.
  \DOIprefix\doi{10.5281/zenodo.10719179}.
\bibitem[{{National Renewable Energy Laboratory
  (NREL)}(2024)}]{NREL2024BuildingTypology}
\bibinfo{author}{{National Renewable Energy Laboratory (NREL)}}.
\newblock \bibinfo{title}{{U.S.} building typology - residential}
  (\bibinfo{year}{2024}).
\newblock \URLprefix
  \url{https://public.tableau.com/app/profile/nrel.buildingstock/viz/USBuildingTypologyResidential/Segments}
  \bibinfo{note}{accessed: 2024-08-23}.
\bibitem[{{Northeast Energy Efficiency Partnerships
  (NEEP)}(2023)}]{NEEP2023ASHP}
\bibinfo{author}{{Northeast Energy Efficiency Partnerships (NEEP)}}.
\newblock \bibinfo{title}{Air source heat pump list}.
\newblock \bibinfo{howpublished}{Northeast Energy Efficiency Partnerships ASHP
  Database} (\bibinfo{year}{2023}).
\newblock \URLprefix \url{https://ashp.neep.org/#!/product_list/}
  \bibinfo{note}{accessed: April 16, 2024}.
\bibitem[{{U.S. Census Bureau}(2023{\natexlab{a}})}]{USCensus2023Vehicles}
\bibinfo{author}{{U.S. Census Bureau}}.
\newblock \bibinfo{title}{Why we ask about... {Vehicles}}
  (\bibinfo{year}{2023}{\natexlab{a}}).
\newblock \URLprefix
  \url{https://www.census.gov/acs/www/about/why-we-ask-each-question/vehicles/}
  \bibinfo{note}{accessed: April 16, 2024}.
\bibitem[{iSe(2023)}]{iSeecars}
\bibinfo{title}{Which vehicle type is the most popular in each state?}
\newblock
  \bibinfo{howpublished}{\url{https://www.iseecars.com/popular-vehicle-type-by-state-study}}
  (\bibinfo{year}{2023}).
\bibitem[{Yuksel and Michalek(2015)}]{yuksel2015effects}
\bibinfo{author}{Yuksel, T.}, and \bibinfo{author}{Michalek, J.~J.}
  (\bibinfo{year}{2015}). \bibinfo{title}{Effects of regional temperature on
  electric vehicle efficiency, range, and emissions in the {{United States}}}.
\newblock \bibinfo{journal}{Environmental science \& technology}
  \emph{\bibinfo{volume}{49}}, \bibinfo{pages}{3974--3980}.
\bibitem[{{U.S. Census Bureau}(2023{\natexlab{b}})}]{USCensus2023Commuting}
\bibinfo{author}{{U.S. Census Bureau}}.
\newblock \bibinfo{title}{Why we ask about... {Commuting}}
  (\bibinfo{year}{2023}{\natexlab{b}}).
\newblock \URLprefix
  \url{https://www.census.gov/acs/www/about/why-we-ask-each-question/commuting/}
  \bibinfo{note}{accessed: April 16, 2024}.
\bibitem[{Res(2024)}]{Resstock}
\bibinfo{title}{{NREL Building Stock Analysis}}.
\newblock
  \bibinfo{howpublished}{\url{https://public.tableau.com/app/profile/nrel.buildingstock/vizzes}}
  (\bibinfo{year}{2024}).
\bibitem[{{American Society of Heating, Refrigerating and Air-Conditioning
  Engineers (ASHRAE)}(2001)}]{ashrae_90_2_2001_addendum}
\bibinfo{author}{{American Society of Heating, Refrigerating and
  Air-Conditioning Engineers (ASHRAE)}}.
\newblock \bibinfo{title}{Ashrae standard 90.2-2001 addendum i}.
\newblock
  \bibinfo{howpublished}{\url{https://www.ashrae.org/File\%20Library/Technical\%20Resources/Standards\%20and\%20Guidelines/Standards\%20Addenda/90-2-2001\_Add-i.pdf}}
  (\bibinfo{year}{2001}).
\newblock \bibinfo{note}{Accessed: 2024-09-06}.
\bibitem[{Ene(????)}]{EnergyStar}
\bibinfo{title}{County level design temperature reference guide}.
\newblock
  \bibinfo{howpublished}{\url{https://www.energystar.gov/partner_resources/residential_new/working/hvac/hvac_designers/design_temp_limits}}
  ().
\bibitem[{Cen(????)}]{Census}
\bibinfo{title}{{United States Census Bureau}}.
\newblock
  \bibinfo{howpublished}{\url{https://www.census.gov/topics/employment/commuting/guidance/commuting.html}}
  ().
\bibitem[{EV(2025)}]{EV}
\bibinfo{title}{{Fuel Economy, U.S. Department of Energy }}.
\newblock
  \bibinfo{howpublished}{\url{https://www.fueleconomy.gov/feg/evtech.shtml}}
  (\bibinfo{year}{2025}).
\bibitem[{ASH(2012)}]{ASHRAE2023Meteo}
\bibinfo{title}{{International Energy Conservation Code} {(IECC)}}
  (\bibinfo{year}{2012}).
\newblock \URLprefix
  \url{https://codes.iccsafe.org/content/IECC2012P5/chapter-3-re-general-requirements}
  \bibinfo{note}{accessed: February 16}.

\end{thebibliography}

\clearpage
\begingroup
\hypersetup{hidelinks}
\listoffigures
\endgroup

\end{document}